\documentclass[prd,nofootinbib,tightenlines]{revtex4}
\def\journal#1, #2, #3#4, #5#6#7#8    {
    {#1~} {#2}  (#5#6#7#8) #3#4}

\usepackage{amssymb}
\usepackage{amsmath}
\usepackage{lscape}
\begin{document}


\renewcommand{\thesection}{\arabic{section}}
\renewcommand{\thesubsection}{\thesection.\arabic{subsection}}
\renewcommand{\theequation}{\arabic{equation}}
\newcommand{\pv}[1]{{-  \hspace {-4.0mm} #1}}

\baselineskip=14pt


\begin{center}
{\bf  \Large $ \kappa$-Minkowski spacetime and the star product realizations}
 
\bigskip

S. Meljanac {\footnote{e-mail: meljanac@irb.hr}}, A. Samsarov
 {\footnote{e-mail: asamsarov@irb.hr}}
and M. Stoji\'c {\footnote{e-mail: marko.stojic@zg.htnet.hr}} \\  
 Rudjer Bo\v{s}kovi\'c Institute, Bijeni\v cka  c.54, HR-10002 Zagreb,
Croatia \\[3mm] 

 Kumar S. Gupta {\footnote{e-mail: kumars.gupta@saha.ac.in}} \\  
 Theory Division, Saha Institute of Nuclear Physics, 1/AF,
 Bidhannagar, Calcutta 700064, India \\[3mm]


\end{center}
\setcounter{page}{1}
\bigskip


\begin{abstract} 
 We investigate a Lie algebra-type $ \; \kappa$-deformed Minkowski
 spacetime with undeformed Lorentz algebra and mutually commutative
 vector-like Dirac derivatives. There are infinitely many realizations
 of $ \; \kappa$-Minkowski space. The coproduct and the star
 product corresponding to each of them are found. An explicit connection
 between realizations and orderings is established and the relation
 between the coproduct and the star product, provided through an exponential
 map, is proved.
 Utilizing the properties
 of the {\em{natural}} realization, we construct a scalar
 field theory on $ \; \kappa$-deformed Minkowski space and show that
 it is equivalent to the scalar, nonlocal, relativistically invariant
 field theory on the ordinary Minkowski space. This result is
 universal and does not depend on the realizations, i.e. orderings used.
\end{abstract}
 

\maketitle



\section{Introduction}
   
    One of the problems whose solution is still being sought involves
    how gravity and quantum field theory can be reconciled.
     The efforts to bring these two theories
    closer have resulted in many new developments in 
     theoretical physics and mathematics.
    In this regard, noncommutative geometry
    appears as a promising framework for unifying gravity and
    quantum field theory. 
    In order to
    accomodate for quantum aspects of spacetime, spacetime
    noncommutativity has been studied extensively
    \cite{1},\cite{2},\cite{3},\cite{bal},\cite{4},\cite{5} and two main
    directions in the investigation have focused on the
    canonical noncommutative spacetime and $ \; \kappa$-Minkowski
    noncommutative spacetime \cite{6}-\cite{28}, respectively. 

    To describe the short-distance (Planck scale) structure of
    spacetime, there appeared a possible necessity for modification
    of the existing symmetries. Specifically, it might require some
    deformation of Poincare symmetry and some arguments would suggest
    that the symmetries of the $ \; \kappa$-deformed Minkowski space
    should be described in terms of a Hopf algebra \cite{7},\cite{9}. For these
    reasons, it is of importance to classify noncommutative (NC) spaces and to
    investigate their properties. In order to make a step forward in
    this direction, we analyze a $ \; \kappa$-deformed Minkowski
    space which is a simple example of a Lie algebra type
    NC space 
    \cite{29}.

   Recently, a new motivation for studying $ \; \kappa$-deformed
   spacetime has appeared within the programme of the so-called doubly
   special relativity (DSR) \cite{15},\cite{16},
    which is a special relativity theory
   extended to include a second invariant parameter $ \; \kappa,$ a
   parameter of mass dimension, besides the velocity of light.
Also, $ \; \kappa$-Minkowski space serves as a playground for
   constructing a field theory on it and then discussing its physical
   properties. 
    The attempts of this type have been undertaken by many
   authors \cite{6}-\cite{28}.

 In $ \; \kappa$-deformed Minkowski space, the noncommutativity of
 coordinates depends on a deformation parameter $ \; a = \frac{1}{\kappa} \;$. Generally,
  this parameter can be considered as an arbitrary
 vector $ \; a \;$ in Minkowski space with all components different
 from zero. In Refs.\cite{22} and \cite{23}, the most general
 deformation of the Euclidean space with the vector 
  $ \; a \;$ in an arbitrary direction has been discussed.
 In the present paper we restrict to the case where only the time
 component of the vector $ \; a \;$ is different from zero.
 The dimensional parameter $ \; a = \frac{1}{\kappa} \; $ has a
 very small value which yields the undeformed Minkowski space in the
 limit $ \; a \rightarrow 0.$ The NC coordinates and the generators of
  rotations form an extended Lie algebra. The 
 subalgebra which includes the rotation generators is the ordinary undeformed
 Lorentz algebra $ \; SO_{a}(1,n-1).$ We assume that Dirac
 derivatives mutually commute and transform as a vector
 representation of the Lorentz algebra. An infinite number of
 realizations \cite{22},\cite{23}
 of $ \; \kappa$-deformed Minkowski space is found in
 terms of commutative coordinates and derivatives. These realizations
 are parametrized by arbitrary invertible functions, with suitably
 imposed conditions
   leading to the undeformed
 Minkowski space in the limit $ \; a \rightarrow 0.$
 Each such
 realization induces an ordering prescription, a coproduct and a star
 product associated to it. For a special choice of the parameter
 functions, some particularly simple realizations are obtained whose
 corresponding star products are known in the literature \cite{22}.
The choice of the particular realization in a way corresponds to a
 particular choice of gauge, so although $ \; \kappa$-Minkowski
 space may be realized in many different ways, physical results do not
 depend on concrete realization, i.e. ordering, as long as all
 commutators between NC coordinates and the Dirac derivatives are fixed. 
In the analyzis of $ \; \kappa$-Minkowski
spaces we use the methods developed for deformed single and
multimode oscillators in the Fock space representations \cite{30}-\cite{36}.
In particular, we use the methods for constructing deformed
creation and annihilation operators in terms of ordinary bosonic
multimode oscillators. Also, we
employ the construction of transition number operators and, in
general, of generators proposed in Refs.\cite{31},\cite{32},\cite{34}.

The crucial point emphasized in this paper is that NC spaces,
particularly the $ \; \kappa$-deformed Minkowski space considered
here, can be realized in many ways in terms of commutative variables
and all these realizations are on equal footing with physical results
not depending on the particular realization taken. 
For fixed deformed Heisenberg algebra,
there is a one-to-one
 correspondence between the realization of NC space and the
ordering prescription and for each such realization there exists
a coproduct and a particular star product corresponding to it.
In a recent paper \cite{26} it was shown that a free scalar field theory
on $ \; \kappa$-deformed Minkowski space can be expressed as a
nonlocal, relativistically invariant field theory on the ordinary Minkowski space. 
The authors of the Ref.\cite{26} defined
 a special star product by using, from the very beginning, a particular
``right'' ordering prescription.
In the present paper we also construct a free scalar field theory on 
$ \; \kappa$-deformed Minkowski space in terms of a particularly
chosen star product, namely the star product corresponding to the
{\em{natural}} realization, where the Dirac derivative
is identified with the ordinary derivative. We show that the corresponding free
scalar field theory on Minkowski space is particularly simple,
nonlocal, relativistically invariant and does not depend on the
realization/ordering taken, i.e. 
does not depend on the particular realization of 4-momenta and
coordinates defining the related NC space.

 This paper is arranged as follows. In section 2.1 we review
  the results of Ref.\cite{22} which are important for the
 investigations carried out in this paper, including the notions of 
 $ \; \kappa$-deformed Euclidean space, its realizations in terms of
 commutative coordinates, and the realizations of the
 undeformed rotation algebra $ \; SO_{a}(n) \;$ compatible with $ \; \kappa$
 deformation. All these realizations are introduced in such a way as to
 retain the structure of the extended Lie algebra. They describe the Lie
 algebra type deformation of space where the rotation
 algebra is undeformed, while the corresponding coalgebra is deformed. The Dirac
 derivative and the invariant Klein Gordon operator, together with
 their realizations, are also introduced in this section. In section
 2.3  we exhibit the one-to-one correspondence between the ordering
 precription and the type of mapping between the NC $ \; \kappa$-deformed
 space and the ordinary undeformed space.
 Here we also give the general BCH formula for the Lie algebra corresponding
 to the $ \; \kappa$ type of deformation of space. In section 2.4 the
 notions of the T- operator and a star product are introduced. Knowing the
 star product in a given realization, the T- operator can be used to
 make a transition to another, equivalent star product.
 This new star
 product corresponds to a realization which is related to first one
 through a similarity transformation by the T- operator.
 In section 3,
 a derivation of the star product formula corresponding to the $ \;
 \varphi$ realization, introduced in section 2.1, is presented. The
 final result is identified with an important relation between the
 coproduct and the star product in terms of an exponential map for a given
 realization. This result can be extended to hold for a
 general Lie algebra. In section 4 all results obtained are
 continued to the Minkowski space with suitable modifications required
  to accomodate for a change in the signature.
  Here we introduce a
 particularly interesting realization, the
 so-called {\em{natural}} realization, discuss its properties and
 derive the star product corresponding to it. We then use this result
 to construct the free scalar field theory on the $ \; \kappa$-deformed
 Minkowski space and to show that it reduces to free, nonlocal scalar
 field theory on the ordinary Minkowski spacetime, no matter which
 realization of the deformed space is used. In constructing the field
 theory on $ \; \kappa$-deformed
 Minkowski space, the emphasis is on the star product corresponding to
 the {\em{natural}} realization because it is the only one with this
 preferable property.  After a short summary, there are two Appendices
 which refer mainly to the derivation of the star product, taking
 place in section 3. 


\section{Algebra of $ \; \kappa $- deformed Euclidean space}

\subsection{The mapping between commutative and noncommutative coordinates}

We consider a noncommutative (NC) space \cite{22} with
coordinates $ \; \hat{x}_{1}, \hat{x}_{2},..., \hat{x}_{n}, \; $
satisfying the following Lie algebra type relations:
\begin{equation}
\lbrack\hat{x}_{\mu},\hat{x}_{\nu}\rbrack=
iC_{\mu\nu\lambda}\hat{x}_{\lambda}
= i(a_{\mu}\hat{x}_{\nu}-a_{\nu}\hat{x}_{\mu}),
\end{equation}
where $ a_{1},a_{2},...,a_{n}$ are  constant real parameters
describing a deformation of Euclidean space. The structure
constants are
\begin{equation}
C_{\mu\nu\lambda}=a_{\mu}\delta_{\nu\lambda}-a_{\nu}\delta_{\mu\lambda}.
\end{equation}
We choose   $ a_{1}=a_{2}=...=a_{n-1}=0, a_{n}=a $ and use  Latin
indices  for the subspace $(1,2,...,n-1)$ and Greek indices  for
the whole space $(1,2,...,n)$. Then the algebra of
the NC coordinates becomes
\begin{equation} \label{com1}
\lbrack\hat{x}_{i},\hat{x}_{j}\rbrack=0, \qquad
\lbrack\hat{x}_{n},\hat{x}_{i}\rbrack=ia\hat{x}_{i},\qquad
i,j=1,2,...,n-1.
\end{equation}
 There exist realizations of the
NC coordinates $\hat{x}_{\mu}$ in terms of ordinary commutative
coordinates  $ x_{1},x_{2},...,x_{n}$ and their derivatives
$\partial_{1},\partial_{2},...,\partial_{n},$  where
$\partial_{\mu}=\frac{\partial}{\partial x_{\mu}}.$ A realization
of the NC coordinates $\hat{x}_{\mu}$ satisfying
the algebra (\ref{com1})  is
\begin{equation} \label{mapping}
\begin{array}{c}
\hat{x}_{i}=x_{i}\varphi(A),\\
\\
\hat{x}_{n}=x_{n}\psi(A)+iax_{k}\partial_{k}\gamma(A),\\
\end{array}
\end{equation}
where  $A=ia\partial_{n}$ and the summation over repeated indices is understood.
It is assumed that $\; \varphi \;$ and $\; \psi \; $ are positive functions 
 obeying the conditions 
\begin{equation}
\varphi(0)=1,\quad  \psi(0)=1,
\end{equation}
The function $ \; \gamma \; $ satisfy
\begin{equation} \label{constraint}
\frac{\varphi'}{\varphi}\psi=\gamma-1,
\end{equation}
where $\varphi'=\frac{d\varphi}{dA}$. It is understood that the
quantity $\gamma(0)=\varphi'(0)+1$ is finite.
In the following discussion we take the vacuum state to be
represented by  $1$, i.e. $\; |0 \rangle \equiv 1, \; $ so that it is
annihilated by all derivatives $\; {\partial}_{\mu} $.

Next we introduce the generators $M_{\mu\nu}, \;\; M_{\mu\nu} =
-M_{\nu\mu} \; $
satisfying the ordinary undeformed $SO_a(n)$ algebra
\begin{equation} \label{com2}
\lbrack M_{\mu\nu},M_{\lambda\rho}\rbrack =
\delta_{\nu\lambda}M_{\mu\rho}-\delta_{\mu\lambda}M_{\nu\rho}
-\delta_{\nu\rho}M_{\mu\lambda}+\delta_{\mu\rho}M_{\nu\lambda},
\end{equation}
or in a more practical form
\begin{equation} \label{com3}
\begin{array}{c}
\lbrack M_{ij},M_{jn}\rbrack=M_{in},\\
\\
\lbrack M_{in},M_{jn}\rbrack=-M_{ij}.
\end{array}
\end{equation}
If we want to extend the algebra to include the action of the
generalized rotation generators $M_{\mu\nu}$ on NC coordinates in
such a way as to keep the Lie algebra structure, then we are brought
to the unique
commutation relations \cite{19},\cite{22} 
\begin{equation} \label{com4}
\lbrack M_{in},\hat{x}_{n}\rbrack=\hat{x}_{i}+ia M_{in},
\end{equation}
\begin{equation} \label{com5}
\lbrack
M_{in},\hat{x}_{j}\rbrack=-\delta_{ij}\hat{x}_{n}+iaM_{ij},
\end{equation}
that, together with (\ref{com1}) and (\ref{com2}),
satisfy the Jacobi identities
$$
\lbrack M_{\alpha\beta},\lbrack
\hat{x}_{\mu},\hat{x}_{\nu}\rbrack\rbrack+ \lbrack
\hat{x}_{\mu},\lbrack \hat{x}_{\nu},M_{\alpha\beta}\rbrack\rbrack+
\lbrack\hat{x}_{\nu},\lbrack
M_{\alpha\beta},\hat{x}_{\mu}\rbrack\rbrack=0,
$$
$$
\lbrack M_{\alpha\beta},\lbrack
M_{\gamma\delta},\hat{x}_{\mu}\rbrack\rbrack+ \lbrack
M_{\gamma\delta},\lbrack
\hat{x}_{\mu},M_{\alpha\beta}\rbrack\rbrack+
\lbrack\hat{x}_{\mu},\lbrack
M_{\alpha\beta},M_{\gamma\delta}\rbrack\rbrack=0.
$$
 These are the necessary and sufficient
conditions for the consistency of the extended Lie algebra with generators
$\hat{x}_{\lambda}$ and $M_{\mu\nu}$.
It may be noted that (\ref{com4}), (\ref{com5}) have a smooth limit
$[M_{\mu\nu},x_{\lambda}]=x_{\mu}\delta_{\nu\lambda}-x_{\nu}\delta_{\mu\lambda},$
when $a_{\mu}\rightarrow 0.$
There are infinitely many representations of  $M_{\mu\nu}$ in
terms of commutative  coordinates $x_{\lambda},$ and their derivatives
$\partial_{\lambda}$. If we require generators  $M_{\mu\nu}$ to be  linear
in $x$ with an infinite series in $\partial$, then there
emerge two families of solutions consistent simultaneously with
the relations (\ref{com3}), (\ref{com4}) and (\ref{com5}). One family is for $ \psi = 1$
 and the other is for $ \psi = 1 + 2A $. Here we focus our attention
 only on the first family which gives us solutions of the form
\begin{equation} \label{trt11}
\begin{array}{c}
M_{ij}=x_{i}\partial_{j}-x_{j}\partial_{i},\\
\\
M_{in}=x_{i}\partial_{n} \varphi\frac{e^{2A}-1}{2A}     
-x_{n}\partial_{i} \frac{1}{\varphi}+ ia
x_{i}\Delta \frac{1}{2\varphi}  -
 ia x_{k}\partial_{k}\partial_{i} \frac{\gamma}{\varphi},\\
\\
M_{ni}=-M_{in},
\end{array}
\end{equation}
where $\Delta=\partial_{k}\partial_{k}, \;\;
 \gamma=\frac{\varphi'}{\varphi}+1 \; $
 and the summation over
repeated indices is understood. This realization can be parametrized by an arbitrary
positive function $\varphi(A),$ \quad $ \varphi(0)=1.$ 
To complete the settings we are dealing with, it is natural to introduce
the Dirac derivatives  $D_{\mu}$ and an invariant operator
 $\Box$ as
\begin{equation} \label{com6}
\begin{array}{c}
\lbrack M_{\mu\nu},D_{\lambda}\rbrack
=\delta_{\nu\lambda}D_{\mu}-\delta_{\mu\lambda}D_{\nu},\\
\\
\lbrack D_{\mu},D_{\nu}\rbrack=0.  \quad  \lbrack M_{\mu\nu}, \Box \rbrack=0,
 \quad  \lbrack \Box, {\hat{x}}_{\mu} \rbrack= 2 D_{\mu}.
\end{array}
\end{equation}
Realizations of the above operators in terms of the commutative
variables have the form
\begin{equation} \label{com7}
\begin{array}{c}
D_{i}=\partial_{i} \frac{e^{-A}}{\varphi},\\
\\
D_{n}=\partial_{n} \frac{\sinh A}{A} + ia \Delta \frac{e^{-A}}{2\varphi^2},
\end{array}
\end{equation}
\begin{equation} \label{com8}
\Box
 = \Delta \frac{e^{-A}}{\varphi^2} - \partial_{n}^2
 \frac{2\lbrack 1- \cosh A \rbrack}{A^2}.
\end{equation}
Similar expressions for $\; M_{\mu \nu}, \; D_{\lambda}, \; $ as given in
Eqs.(\ref{trt11})-(\ref{com8}), have been obtained in
Refs.\cite{19},\cite{21}, but only for three particular choices of the
function $\; \varphi, \; $ namely $\; \varphi = 1, \; \varphi =
e^{-A} \; $ and $\; \varphi = \frac{A}{e^A - 1}. $
Now we calculate the commutation relations between the NC coordinates
$\hat{x}_{\mu}$ and the Dirac derivatives $D_{\nu}$:
\begin{equation} \label{com11}
\begin{array}{c}
\lbrack
D_{i},\hat{x}_{j}\rbrack=\delta_{ij}(-iaD_{n}+\sqrt{1-a^2D_{\mu}D_{\mu}}),\\
\\

\lbrack D_{i},\hat{x}_{n}\rbrack=0 , \\
\\
\lbrack D_{n},\hat{x}_{i}\rbrack=iaD_{i},\\
\\
\lbrack D_{n},\hat{x}_{n}\rbrack=\sqrt{1-a^2D_{\mu}D_{\mu}}.
\end{array}
\end{equation}
These relations are fixed by all previous commutation relations
(\ref{com1}),(\ref{com3}),(\ref{com4}),(\ref{com5}),(\ref{com6})
 and they involve only the deformation parameter $a$.


\subsection{Lebniz rules and the coproducts}

An important ingredient of the symmetry structure of
$\kappa$-deformed  space are the Leibniz  rules of generators of
rotations and derivatives. For the case in consideration, the Leibniz
rule for the rotation generators looks like
$$
M_{ij}(\hat{f} \cdot \hat{g})=(M_{ij} \hat{f}) \cdot \hat{g} + \hat{f}
\cdot (M_{ij} \hat{g}),\\
$$
$$
M_{in}(\hat{f} \cdot \hat{g})=(M_{in} \hat{f}) \cdot
 \hat{g} + (e^A \hat{f}) \cdot (M_{in}
 \hat{g})+ia(\partial_j\frac{1}{\varphi(A)} \hat{f}) \cdot (M_{ij} \hat{g}),
$$
where $ \hat{f} $, $ \hat{g} $ are functions of the NC coordinates $\hat{x}_{\mu}$.
Similarly, we have the expressions 
$$
D_n (\hat{f} \cdot \hat{g})= (D_n \hat{f}) \cdot (e^{-A} \hat{g}) +
 \Big(\frac{iaD_n+\sqrt{1-a^2D_{\mu}D_{\mu}}}{1-a^2D_k D_k} \hat{f} \Big)
\cdot (D_n \hat{g})
$$
$$
 + \Big( iaD_i\frac{iaD_n+\sqrt{1-a^2D_{\mu}D_{\mu}}}{1-a^2D_kD_k} \hat{f} \Big)
\cdot ( D_i \hat{g}),
$$
$$
 D_i (\hat{f} \cdot \hat{g}) = (D_i \hat{f}) \cdot 
  (e^{-A} \hat{g}) + \hat{f} \cdot ( D_i \hat{g}),
$$
as the Leibniz rules for the Dirac derivative.

The above equations for the Leibniz rules comprise the notion of
coproducts. In the case of the rotation generators, the coproduct looks as
$$
\triangle M_{ij}=M_{ij}\otimes 1+1\otimes M_{ij},
$$
$$
\triangle M_{in}=M_{in}\otimes 1+e^A\otimes M_{in}+iaD_je^A\otimes
M_{ij},
$$
while, in the case of the Dirac derivative, it is given by
$$
\triangle
D_n=D_n\otimes\Big(-iaD_n+\sqrt{1-a^2D_{\mu}D_{\mu}}\Big)+
\frac{iaD_n+\sqrt{1-a^2D_{\mu}D_{\mu}}}{1-a^2D_k D_k}\otimes D_n
$$
$$
 +iaD_i\frac{iaD_n+\sqrt{1-a^2D_{\mu}D_{\mu}}}{1-a^2D_kD_k}\otimes D_i,
$$
\begin{equation} \label{com9}
\triangle
 D_i=D_i\otimes\Big(-iaD_n+\sqrt{1-a^2D_{\mu}D_{\mu}}\Big)+1\otimes D_i,
\end{equation}
where $e^A$ is defined as
$$
e^{-A}=-iaD_{n}+\sqrt{1-a^2D_{\mu}D_{\mu}}.
$$
The final result for the coproduct
depends only on $aD_{n}$, and $a^2D_{\mu}D_{\mu}$. In the limit
$a\rightarrow 0,$  it gives the ordinary undeformed coproduct for
$M_{\mu\nu},$ i.e. $D_{\mu},$ respectively.
The notion of the coproduct leads to the observation
that the generators of the $\kappa$-deformed symmetry are elements
of a Hopf algebra.
The coproduct $\triangle$, which we have  determined for $M_{\mu\nu}$ and
$D_{\mu},$  extends multiplicatively to the whole algebra $ISO_a(n)$, which becomes 
a Hopf algebra in this way.  Some examples of the Poincar\'{e} invariant
interpretation of NC spaces and of the twisted Poincar{\'e}
coalgebra were also considered in \cite{37},\cite{38}.

   We note that from Eqs.(\ref{mapping}) it follows that
$$
 [ {\partial}_{i}, {\hat{x}}_{j}] = {\delta}_{ij} \varphi (A);   \; \quad
  [ {\partial}_{i}, {\hat{x}}_{n}] = i a {\partial}_{i} \gamma (A)
$$
\begin{equation} \label{first1natural}
 [ {\partial}_{n}, {\hat{x}}_{i}] = 0;   \; \quad
  [ {\partial}_{n}, {\hat{x}}_{n}] = 1.
\end{equation}
The corresponding coproducts $ \; {\triangle}_{\varphi} \; $ are
$$
 {\triangle}_{\varphi} ( {\partial}_{n}) = {\partial}_{n} \otimes 1 +
 1 \otimes {\partial}_{n} = {\partial}_{n}^{x} + {\partial}_{n}^{y},
$$
\begin{equation} \label{partialcoproduct}
 {\triangle}_{\varphi} ( {\partial}_{i}) = {\partial}_{i}^{x} 
    \frac{\varphi ( A_{x} + A_{y})}{\varphi (A_{x})} + {\partial}_{i}^{y}
   \frac{\varphi ( A_{x} + A_{y})}{\varphi (A_{y})} e^{A_{x}},
\end{equation}
where $ \; A_{x} = ia {\partial}_{n}^{x}  \; $ and similarly for $ \; A_{y}.$ 
They are consistent with Eqs.(\ref{com9}) and (\ref{com7}).
At this point it is convenient to introduce a notion of {\em{natural}}
realization, namely the $ \; \varphi $ realizations (\ref{mapping})
are natural with respect to Eqs.(\ref{first1natural}).
The {\em{natural}} realization is generally characterized by the
reqirement $ \; D_{\mu} \equiv {\partial}_{\mu}. $

 Particularly, for the deformed Heisenberg algebra (\ref{com11}),
 which is fixed in the case of the extended Lie algebra introduced in
 section 2.1, the {\em{natural}} realization is given by
\begin{equation} \label{natrellie}
  \hat{x}_{\mu}^{Nat}  = x_{\mu} \sqrt{1 - a^2 {\partial}^{2}} - i
  C_{\alpha \mu \nu} {\partial}_{\alpha},
\end{equation}
with $ \; C_{\alpha \mu \nu} {\partial}_{\alpha} = a_{\alpha}
{\delta}_{\mu \nu} - a_{\mu} {\delta}_{\alpha \nu}. $

\subsection{The correspondence of mapping and ordering}

In this section we want to establish the relation between different
realizations of NC space and different ordering prescriptions. In
other words, we want to make clear that for
each mapping of the form (\ref{mapping}) there exists an accompanying ordering
prescripton. In the previous section we have introduced the
{\em{natural}} realization which is seen as to not belong to the family
(\ref{mapping}). For this type of realization of NC space there also
exists one particular ordering prescription.
In this section we shall establish the general form of the orderings
 corresponding to the family of realizations
(\ref{mapping}), while in the section 4.2, we shall deduce the form of the
ordering prescription which belongs to the {\em{natural}} realization.
  To begin with, let us define the "vacuum" state
\begin{equation}
\begin{array}{c}
|0>=1, \qquad D_{\mu}|0>=\partial_{\mu}|0>=0,
\end{array}
\end{equation}
then for a given realization of NC space, described by $ \varphi (A)$, there is a
unique mapping from the set of fields $f(x)$ of the commutative coordinates $x_{\mu}$
 to the set of fields $ \hat{f}(\hat{x}_{\mu})$ of the
NC coordinates $\hat{x}_{\mu}$. This mapping $\; {\Omega}_{\varphi} \;$ can be characterized as
\begin{equation}
 {{\Omega}_{\varphi}}^{-1}: \hat{f}(\hat{x}) \longrightarrow f_{\varphi}(x) \quad
\mbox{such that}  \quad  \hat{f}(\hat{x}_\varphi)|0> = f_{\varphi}(x).
\end{equation}
  The fields or functions on NC space are conveniently introduced
  through the Fourier transform
\begin{equation} \label{Fouriertransform}
 \hat{f}(\hat{x}) = \int d^{n} x \; {\tilde{f}}_{\varphi}(k)
 :e^{ik_{\alpha} {\hat{x}}_{\alpha}}:_{\varphi},
\end{equation}
where  $ \; {: :}_{\varphi} \; $ denotes the ordering that
corresponds to the realization (\ref{mapping}) of NC space, parametrized by the
function  $ \; \varphi. \; $ The way in which this correspondence is
realized will be made clear from the following consideration. 
Thus, to make a connection between the particular realization of NC space
and the corresponding ordering prescription, we impose the 
requirement on the ordered exponentials
\begin{equation} \label{connectionbetweenorderingandmapping}
 :e^{ik \cdot {\hat{x}}_{\varphi}}:_{\varphi} |0 \rangle
 = e^{ik \cdot x},
\end{equation}
appearing in the expansion (\ref{Fouriertransform}).
For example, we have the following expressions for the left, right
 and symmetric ordering prescriptions, respectively:
$$
 :e^{ik_{\alpha} {\hat{x}}_{\alpha}}:_{L} \equiv e^{ik_{n} {\hat{x}}_{n}}
        e^{ik_{i} {\hat{x}}_{i}} =
 e^{ik_{n} {\hat{x}}_{n} + ik_{i} {\hat{x}}_{i} {\varphi}_{S}(-ak_{n})
 e^{- a k_{n}}}, 
$$
$$
 :e^{ik_{\alpha} {\hat{x}}_{\alpha}}:_{R} \equiv 
        e^{ik_{i} {\hat{x}}_{i}} e^{ik_{n} {\hat{x}}_{n}} =
 e^{ik_{n} {\hat{x}}_{n} + ik_{i} {\hat{x}}_{i} {\varphi}_{S}(-ak_{n})},
$$
\begin{equation} \label{specialordering}
 :e^{ik_{\alpha} {\hat{x}}_{\alpha}}:_{S} = e^{ik_{\alpha} {\hat{x}}_{\alpha}}.
\end{equation}
In the above relations we have introduced the function
\begin{equation} \label{symmetric}
 {\varphi}_{S}(A) = \frac{A}{e^{A} - 1}, 
\end{equation}
where $ \; A = ia {\partial}_{n} \equiv -a k_{n}.  \; $
Let us first consider the left ordered Fourier exponential.
 What we are interested here is to find the realization
(\ref{mapping})
for which this left ordered exponential will satisfy the
 requirement (\ref{connectionbetweenorderingandmapping}).
Because all realizations are of the form (\ref{mapping}), this 
basicaly reduces to the problem of finding the correct function $ \; \varphi. \; $
It can be easily shown that for the left ordering this function is
$ \; \varphi = e^{-A}. \; $ By taking the same reasoning for the right
 ordered Fourier exponential, we end up with the function $ \; \varphi = 1 \; $
as the one that defines the mapping (\ref{mapping}) for which the
 right ordered exponential satisfies the requirement
 (\ref{connectionbetweenorderingandmapping}). In exactly the same
 way, the symmetric ordering is specified by the function
$ \; \varphi = {\varphi}_{S}.$
There is also a family of ordering prescriptions, interpolating
between left and right, namely,
\begin{equation} \label{amelino}
 :e^{ik_{\alpha} {\hat{x}}_{\alpha}}:_{C} = e^{i \lambda k_{n} {\hat{x}}_{n}} 
            e^{ik_{i} {\hat{x}}_{i}} e^{i (1 - \lambda)k_{n} {\hat{x}}_{n}},
\end{equation}
which corresponds to the realization characterized by the function
 $ \; \varphi = e^{- \lambda A}. \; $ For $ \; \lambda = \frac{1}{2}, \; $ 
 this ordering prescription coincides with the one used in \cite{25}.
Generally, for the ordering prescription corresponding to the general mapping 
(\ref{mapping}) with the function $ \; \varphi, \; $ we have
\begin{equation} \label{ordering}
 :e^{ik_{\alpha} {\hat{x}}_{\alpha}}:_{\varphi}  =
 e^{ik_{n} {\hat{x}}_{n} + ik_{i} {\hat{x}}_{i} \frac{{\varphi}_{S}(-ak_{n})}
 {\varphi(-ak_{n})}}. 
\end{equation}
It is clear that (\ref{ordering}) incorporates all special cases
(\ref{specialordering}) for the left, right and symmetric ordering,
as well for the case (\ref{amelino}), respectively.

Thus, the above consideration has shown that for each mapping, here specified by the 
function $ \; \varphi, \; $
there exists a particular basis $ \; :e^{ik \cdot {\hat{x}}_{\varphi}}:_{\varphi} \; $ 
of the ordered Fourier exponentials. That is, for each mapping there is a particular
ordering prescription with the mutual correspondence between these two
notions provided through the requirement 
(\ref{connectionbetweenorderingandmapping}). 
The aforementioned correspondence can be looked upon from a slightly different
point of view. In respect to this, one can consider a more general form of
the relation (\ref{connectionbetweenorderingandmapping}), namely,
\begin{equation} \label{generalconnectionbetweenorderingandmapping}
 :e^{ik \cdot {\hat{x}}_{\varphi}}:_{\chi} |0 \rangle
 = e^{i K^{\chi}_{\varphi} (k) \cdot x},
\end{equation}
where $ \;  K^{\chi}_{\varphi} (k) \; $ is the function defined in 
(\ref{Kmapping}) and has the momentum $ \; k \; $ 
as an argument.
In this expression, one can change realization of NC space as well as
ordering prescription simply by changing the functions  $ \; \varphi, \; \chi, \; $
 respectively. By doing this, the requirement
(\ref{connectionbetweenorderingandmapping}) will be fulfilled only after the changes in
realization and ordering meet each other, i.e. it will be satisfied
only when $ \; \varphi = \chi. \; $
 
The relation (\ref{ordering}) can be deduced from the BCH
formula for the $ \; \kappa$-deformed Lie algebra type of NC space
\begin{equation} \label{BCH}
 e^{ik_{\alpha} {\hat{x}}_{\alpha}} e^{iq_{\alpha} {\hat{x}}_{\alpha}}
 = e^{i {\mathcal{D}}_{\alpha}(k,q) {\hat{x}}_{\alpha}},
\end{equation}
where
$$
 {\mathcal{D}}_{i}(k,q) = k_{i} \frac{{\varphi}_{S}(-ak_{n} -
 aq_{n})}{{\varphi}_{S}(-ak_{n})}
   + q_{i} \frac{{\varphi}_{S}(-ak_{n} -
 aq_{n})}{{\varphi}_{S}(-aq_{n})} e^{- a k_{n}},
$$
\begin{equation} \label{BCH1}
 {\mathcal{D}}_{n}(k,q) = k_{n} + q_{n},
\end{equation}
and from the defining relation for the generalized momentum addition rule
 $ \; k \oplus q \; $ which reads as
\begin{equation} \label{BCHordered}
 :e^{ik_{\alpha} {\hat{x}}_{\alpha}}:_{\varphi} :e^{iq_{\alpha} {\hat{x}}_{\alpha}}:_{\varphi}
 = :e^{i {(k \oplus q)}_{\alpha} {\hat{x}}_{\alpha}}:_{\varphi}.
\end{equation} 
Inserting the relation (\ref{ordering}) into the relation (\ref{BCHordered})
and using (\ref{BCH}), (\ref{BCH1}), we get
$$
 {(k \oplus q)}_{i} = k_{i} \frac{\varphi(-ak_{n} -
 aq_{n})}{\varphi(-ak_{n})}
   + q_{i} \frac{\varphi(-ak_{n} -
 aq_{n})}{\varphi(-aq_{n})} e^{- a k_{n}},
$$
\begin{equation} \label{BCHstar}
 {(k \oplus q)}_{n} = k_{n} + q_{n}.
\end{equation}
We see that the generalized momentum addition rule (\ref{BCHstar})
is fully in agreement with the coproduct $ \: {\triangle}_{\varphi}({\partial}_{\mu}), \; $
obtained in the previous section, relation (\ref{partialcoproduct}).
Precisely, if we set $ \: k = -i {\partial}_{1}, \;\;  q = -i {\partial}_{2}, \; $
then (\ref{partialcoproduct}) and (\ref{BCHstar})
 can be related as $ \: {\triangle}_{\varphi}(\partial) = 
   {\partial}_{1} \oplus {\partial}_{2}. $


\subsection{T- operator}

Now we introduce a  star product in a given $\varphi$-realization  in the
following way. As in the previous section,
 we define the "vacuum" state as
\begin{equation}
\begin{array}{c}
|0>=1, \qquad D_{\mu}|0>=\partial_{\mu}|0>=0.
\end{array}
\end{equation}
and introduce the unique map $\; {\Omega}_{\varphi} \;$ which,
for a given realization described by $\varphi(A),$ maps
 the commutative algebra of functions $f(x)$ 
into the noncommutative algebra of functions $ \hat{f}(\hat{x}_{\mu}).$
The map $\; {\Omega}_{\varphi} \;$  can be uniquely characterized by
\begin{equation}
 {\Omega}_{\varphi}: f(x) \longrightarrow \hat{f}({\hat{x}}_{\varphi}) \quad
\mbox{such that}  \quad  \hat{f}(\hat{x}_\varphi)|0> = f(x).
\end{equation}
If we now have two functions  $ \hat{f}(\hat{x}),  \hat{g}(\hat{x})$ of NC coordinates, make their
product $ \hat{f}(\hat{x}) \cdot  \hat{g}(\hat{x})$ and ask
 which function of commutative coordinates this
combination belongs to (through the mapping $\; {\Omega}_{\varphi} \;$), we arrive at
\begin{equation} \label{starproduct}
(f \star_\varphi g)(x) =
 \hat{f}(\hat{x}_\varphi) \hat{g}(\hat{x}_\varphi) |0> =
 \hat{f}(\hat{x}_\varphi) g(x).
\end{equation}
By this expression the star product in a given $\varphi$ realization is defined.
This product obviously depends on the realization $\varphi$. 
Generally, the star products belonging to two different realizations
$ \; \varphi_{1} \;$ and $ \; \varphi_{2} \;$ can be related to each
other by the so-called T- operator that identifies a class of
equivalent star products. Thus, the T- operator can be characterized
by (see for example Ref.\cite{21})
\begin{equation} \label{tproduct}
 T(f(x) {\star}_{{\varphi}_{1}} g(x)) =
 T(f(x)) {\star}_{{\varphi}_{2}} T(g(x)).
\end{equation}
In Ref.\cite{21} T operator has been obtained by relating the cases
for  $ \; \varphi = e^{-A} \;$  and $ \; \varphi = 1 \;$ to that for $
\; \varphi = \frac{A}{e^{A} -1}, \;$ corresponding to left, right
and symmetric ordering, respectively.
To deduce its explicit form for the case of $ \; \kappa$-deformed
spaces, we take advantage of the following identities:
\begin{equation} \label{identities}
\begin{array}{c}
x_{i}\star_\varphi
f(x)=(\hat{x}_\varphi)_{i}f(x)=x_{i}\varphi(A)f(x),\\
\\
x_{n}\star_\varphi f(x) = (\hat{x}_\varphi)_{n}f(x)=[x_{n} + iax_{k}\partial_{k}\gamma(A)]f(x),
\end{array}
\end{equation}
together with the obvious property $ \; T(x_{i}) = x_{i} \; $ and
$ \; T(x_{n}) = x_{n}. \; $
We can now proceed as follows:
\begin{equation}
 T(x_{i} {\star}_{{\varphi}_{1}} f(x)) =
 T(x_{i}) {\star}_{{\varphi}_{2}} T(f(x)).
\end{equation}
By using (\ref{identities}), we have
\begin{equation} 
 T(x_{i} {\varphi}_{1}(A) f(x)) =
 x_{i} {\varphi}_{2}(A) T(f(x)),
\end{equation}
implying
\begin{equation} \label{tevaluation1}
 \left ( x_{i}T + \frac{\partial T}{\partial {\partial}_{i}} \right )
 {\varphi}_{1}(A) f(x) = x_{i} {\varphi}_{2}(A) T f(x).
\end{equation}
Since (\ref{tevaluation1}) holds for any function $ \; f(x), \; $ it
reduces to
\begin{equation} \label{tevaluation12}
  x_{i} ({\varphi}_{1}(A) - {\varphi}_{2}(A)) T = 
- \frac{\partial T}{\partial {\partial}_{i}} {\varphi}_{1}(A),
\end{equation}
which integrates to 
\begin{equation} \label{toperator}
  T = \lim_{y \rightarrow x} e^{x_{i} {\partial}_{i}^{y} \left (
  \frac{{\varphi}_{2}(A)}{{\varphi}_{1}(A)} - 1 \right )}
 \equiv
   : e^{x_{i} {\partial}_{i} \left (
  \frac{{\varphi}_{2}(A)}{{\varphi}_{1}(A)} - 1 \right )}:,
\end{equation}
 where the symbol :: has the meaning of the usual normal
ordering with
the variables $ \; {x}_{i} \; $ coming to the left with respect to the variables
$ \; {\partial}_{i} \; $ and should not be confused with the $ \; {::}_{\varphi} \; $
used before.
The solution (\ref{toperator}) also satisfies the equation
\begin{equation} \label{tevaluation13}
 \frac{\partial T}{\partial {\partial}_{n}} = i a x_{k} {\partial}_{k} 
 ({\gamma}_{2}(A) - {\gamma}_{1}(A)) T,
\end{equation}
stemming from the condition
\begin{equation}
 T(x_{n} {\star}_{{\varphi}_{1}} f(x)) =
 T(x_{n}) {\star}_{{\varphi}_{2}} T(f(x)).
\end{equation}
Using the formula (\ref{B1}) from Appendix A, we can easily find the inverse of
 (\ref{toperator}) which reads
\begin{equation} \label{invtoperator}
  T^{-1} =
   : e^{x_{i} {\partial}_{i} \left (
  \frac{{\varphi}_{1}(A)}{{\varphi}_{2}(A)} - 1 \right )}:.
\end{equation}
If we introduce the notation 
$ \; T_{1,2} \equiv T  = : e^{x_{i} {\partial}_{i} \left (
  \frac{{\varphi}_{1}(A)}{{\varphi}_{2}(A)} - 1 \right )}:, \; $  we can summarize the properties of the 
$ \; T-\; $ operator as follows:
\begin{enumerate}
\item inverse operator,
\begin{equation} 
  T_{1,2}^{-1} = T_{2,1},
  \end{equation}  
\item transitivity,
 \begin{equation} 
   T_{1,2} T_{2,3}  = T_{1,3}.
  \end{equation}  
 \end{enumerate} 

Finally, using Eq.(\ref{C7}) in Appendix A, we can write the operator T,
 Eq.(\ref{toperator}), as
\begin{equation} \label{toperatoralternativa}
  T =
    e^{x_{i} {\partial}_{i} \ln \frac{{\varphi}_{2}}{{\varphi}_{1}}}.
\end{equation}
It is important to note that we can represent $ \; \hat{x} \; $
 in terms of two representations $ \; {\varphi}_{1}(A)\; $
 and $ \; {\varphi}_{2}(A) \; $ as
\begin{equation} \label{mapping11}
\begin{array}{c}
\hat{x}_{i}={(x_{1})}_{i} {\varphi}_{1}(A) = {(x_{2})}_{i} {\varphi}_{2}(A),\\
\\
\hat{x}_{n}={(x_{1})}_{n} +
ia{(x_{1})}_{k}{(\partial_{1})}_{k}{\gamma}_{1}(A) =
  {(x_{2})}_{n} + ia{(x_{2})}_{k}{(\partial_{2})}_{k}{\gamma}_{2}(A),
\end{array}
\end{equation}
where for the transformation of the coordinates we have
\begin{equation} \label{trans1}
  {(x_{2})}_{i} = T^{-1} {(x_{1})}_{i} T = {(x_{1})}_{i} \frac{{\varphi}_{1}(A)}{{\varphi}_{2}(A)},
\end{equation}
\begin{equation} \label{trans2}
  {(x_{2})}_{n} = T^{-1} {(x_{1})}_{n} T = 
  {(x_{1})}_{n} + ia{(x_{1})}_{k}{(\partial_{1})}_{k}({\gamma}_{1}(A)
  - {\gamma}_{2}(A)).
\end{equation}
Similarly, for the transformation of the derivatives, it follows
\begin{equation} \label{trans3}
  {({\partial}_{2})}_{i} = T^{-1} {({\partial}_{1})}_{i} T =
     {({\partial}_{1})}_{i} \frac{{\varphi}_{2}(A)}{{\varphi}_{1}(A)},
\end{equation}
\begin{equation} \label{trans4}
  {({\partial}_{2})}_{n} = T^{-1} {({\partial}_{1})}_{n} T =
     {({\partial}_{1})}_{n}. 
\end{equation}
Generally, one can introduce a mapping
 $ \; K_{{\varphi}_{2}}^{{\varphi}_{1}}: R^{n} \longrightarrow R^{n}, \; $
transforming the vector $ \; {\partial}_{{\varphi}_{1}} \; $ into
 vector $ \; {\partial}_{{\varphi}_{2}}, \; $
\begin{equation} \label{Kmapping}
  {\partial}_{{\varphi}_{2}} = T^{-1} {\partial}_{{\varphi}_{1}} T =
     K_{{\varphi}_{2}}^{{\varphi}_{1}}({\partial}_{{\varphi}_{1}}). 
\end{equation}
Because of the transitivity property of the $ \; T- \;$operator, the
composition of $ \; K \;$ functions
satisfies the relation
\begin{equation} \label{Kmappingproperty}
   K_{{\varphi}_{2}}^{{\varphi}_{1}}
   K_{{\varphi}_{3}}^{{\varphi}_{2}} =  K_{{\varphi}_{3}}^{{\varphi}_{1}}. 
\end{equation}


\section{Derivation of the star product corresponding to the
  arbitrary ordering prescripton / mapping}

  We now proceed in some detail to find the star product for $ \;
  \kappa$-deformed space, corresponding to the particular
ordering $ \; ::_{\varphi} \; $. It can be deduced from the defining
  relation (\ref{tproduct}), starting from the expression for the left star product,
which is known from the literature as
\begin{equation} \label{leftproduct}
 f(x) \; {\star}_{L} \; g(x) =
  \lim_{\substack{y \rightarrow x\\ z \rightarrow x}} e^{x_{i} {\partial}_{i}^{y}
  (e^{-ia{\partial }_{n}^{z}} - 1)} f(y)g(z) =
   \lim_{y \rightarrow x} e^{-ia x_{i} {\partial}_{i}^{x} {\partial}_{n}^{y}}
   f(x)g(y).
\end{equation}
Thus, from $ \; T(f(x) \;  {\star}_{L} \; g(x)) =
 T(f(x)) \;  {\star}_{\varphi} \;  T(g(x)), \; $ it follows
\begin{equation} \label{d1}
 (\tilde{f} \; {\star}_{\varphi} \; \tilde{g})(x) =
  T(T^{-1} \tilde{f} \; {\star}_{L} \; T^{-1} \tilde{g} ),
\end{equation}
where $ \; \tilde{f} = T f \; $ is introduced. For simplicity, in due course, we drop the symbol
$ \; \sim \; $ from the functions and by using the relations
(\ref{toperator}), (\ref{invtoperator}) and (\ref{leftproduct}), we
arrive at
$$
 (f \; {\star}_{\varphi} \; g)(x) =
   \lim_{w \rightarrow x} e^{x_{j} {\partial}_{j}^{w} \left (
  \varphi (A_{w}) e^{A_{w}} - 1 \right )}
   \lim_{\substack{z \rightarrow w  \\ y \rightarrow w }} e^{w_{j} {\partial}_{j}^{z}
  (e^{- A_{y}} - 1)} \times
$$
\begin{equation} \label{d2}
 \times  \lim_{u \rightarrow z} e^{z_{j} {\partial}_{j}^{u} \left (
  \frac{e^{-A_{u}}}{\varphi (A_{u})}  - 1 \right )}
   \lim_{t \rightarrow y} e^{y_{j} {\partial}_{j}^{t} \left (
  \frac{e^{-A_{t}}}{\varphi (A_{t})}  - 1 \right )} f(u) g(t), 
\end{equation}
with the abbreviation 
$ \; A_{x}=ia\frac{\partial}{\partial x_{n}}, \; $ etc.
The second and the third term in (\ref{d2}) can be contracted with the help of Eq. (\ref{B1}) 
to yield
$$
 (f \; {\star}_{\varphi} \; g)(x) =
   \lim_{w \rightarrow x} e^{x_{j} {\partial}_{j}^{w} \left (
  \varphi (A_{w}) e^{A_{w}} - 1 \right )}
 :e^{z_{j} {\partial}_{j}^{z} (e^{- A_{y}} - 1)}:
   :e^{z_{j} {\partial}_{j}^{z} \left (
  \frac{e^{-A_{u}}}{\varphi (A_{u})}  - 1 \right )}:
$$
\begin{equation} \label{d3}
   \lim_{t \rightarrow y} e^{y_{j} {\partial}_{j}^{t} \left (
  \frac{e^{-A_{t}}}{\varphi (A_{t})}  - 1 \right )} f(u) g(t)
\end{equation}
$$
  = \lim_{w \rightarrow x} e^{x_{j} {\partial}_{j}^{w} \left (
  \varphi (A_{w}) e^{A_{w}} - 1 \right )}
   :e^{N \left ( e^{-A_{y}}
  \frac{e^{-A_{u}}}{\varphi (A_{u})}  - 1 \right )}:
   \lim_{t \rightarrow y} e^{y_{j} {\partial}_{j}^{t} \left (
  \frac{e^{-A_{t}}}{\varphi (A_{t})}  - 1 \right )} f(u) g(t) 
$$
\begin{equation} \label{d4}
  = \lim_{w \rightarrow x} e^{x_{j} {\partial}_{j}^{w} \left (
  \varphi (A_{w}) e^{A_{w}} - 1 \right )}
      \lim_{\substack{u \rightarrow w  \\ y \rightarrow w }} 
     e^{w_{j} {\partial}_{j}^{u} \left ( e^{-A_{y}}
  \frac{e^{-A_{u}}}{\varphi (A_{u})}  - 1 \right )}
   \lim_{t \rightarrow y} e^{y_{j} {\partial}_{j}^{t} \left (
  \frac{e^{-A_{t}}}{\varphi (A_{t})}  - 1 \right )} f(u) g(t). 
\end{equation}
If the obvious relation $ \; \varphi (A_{w}) \lim_{t \rightarrow
 y} f(t) = \lim_{t \rightarrow y} \varphi (A_{t}) f(t) \; $ and
the similar one (\ref{F1}) derived in Appendix B are applied to the
 second and the third factor in (\ref{d4}), we get
$$
 (f \; {\star}_{\varphi} \; g)(x)
  = \lim_{w \rightarrow x} e^{x_{j} {\partial}_{j}^{w} \left (
  \varphi (A_{w}) e^{A_{w}} - 1 \right )}
   \lim_{\substack{u \rightarrow w  \\ t \rightarrow w }} 
     e^{w_{j} {\partial}_{j}^{u} \left ( e^{-A_{t}}
  \frac{e^{-A_{u}}}{\varphi (A_{u})}  - 1 \right )}
    e^{w_{j} {\partial}_{j}^{t} \left (
  \frac{e^{-A_{t}}}{\varphi (A_{t})}  - 1 \right )} f(u) g(t)
$$
\begin{equation} \label{d5}
  = \lim_{w \rightarrow x} \lim_{\substack{u \rightarrow w  \\ t \rightarrow w }}
 e^{x_{j} ({\partial}_{j}^{u} + {\partial}_{j}^{t}) \left (
  \varphi (A_{u} + A_{t}) e^{A_{u} + A_{t}} - 1 \right )}
     e^{w_{j} {\partial}_{j}^{u} \left ( e^{-A_{t}}
  \frac{e^{-A_{u}}}{\varphi (A_{u})}  - 1 \right )}
   e^{w_{j} {\partial}_{j}^{t} \left (
  \frac{e^{-A_{t}}}{\varphi (A_{t})}  - 1 \right )} f(u) g(t). 
\end{equation}
Here we can again use the equivalent representation of the factors
in (\ref{d5}) through the normal ordering (not to be confused with 
$ \; ::_{\varphi}) \; $ and apply (\ref{B1}) to obtain
$$
 (f \; {\star}_{\varphi} \; g)(x)
 = \bigg ( \lim_{\substack{u \rightarrow x  \\ t \rightarrow x }}
    e^{x_{j} {\partial}_{j}^{u} \left (
   \varphi (A_{u} + A_{t}) e^{A_{u} + A_{t}} - 1 \right )}
      e^{x_{j} {\partial}_{j}^{u} \left ( e^{-A_{t}}
  \frac{e^{-A_{u}}}{\varphi (A_{u})}  - 1 \right )} \bigg ) 
  \bigg ( \lim_{\substack{u \rightarrow x  \\ t \rightarrow x }}
   e^{x_{j} {\partial}_{j}^{t} \left (
   \varphi (A_{u} + A_{t}) e^{A_{u} + A_{t}} - 1 \right )}
    e^{x_{j} {\partial}_{j}^{t} \left (
  \frac{e^{-A_{t}}}{\varphi (A_{t})}  - 1 \right )} \bigg ) f(u) g(t)
$$
$$
  = \bigg ( :e^{N \left (
   \varphi (A_{u} + A_{t}) e^{A_{u} + A_{t}} - 1 \right )}:
     :e^{N \left ( e^{-A_{t}}
  \frac{e^{-A_{u}}}{\varphi (A_{u})}  - 1 \right )}: \bigg ) 
 \bigg (  :e^{N \left (
   \varphi (A_{u} + A_{t}) e^{A_{u} + A_{t}} - 1 \right )}:
    :e^{N \left (
  \frac{e^{-A_{t}}}{\varphi (A_{t})}  - 1 \right )}: \bigg ) f(u) g(t)
$$
$$
 = :e^{N \left ( \varphi (A_{u} + A_{t}) e^{A_{u} + A_{t}} e^{-A_{t}}
  \frac{e^{-A_{u}}}{\varphi (A_{u})} - 1 \right )}:
   :e^{N \left (
    \varphi (A_{u} + A_{t}) e^{A_{u} + A_{t}} \frac{e^{-A_{t}}}{\varphi (A_{t})} - 1 \right )}:
    f(u) g(t)
$$
\begin{equation} \label{d6}
  = :e^{N \left (
   \frac{ \varphi (A_{u} + A_{t}) }{\varphi (A_{u})} - 1 \right )}:
   :e^{N \left (
   \frac{ \varphi (A_{u} + A_{t}) }{\varphi (A_{t})} e^{A_{u}} - 1 \right )}:
    f(u) g(t)  
    = \lim_{u \rightarrow x} e^{x_{j} {\partial}_{j}^{u} \left (
  \frac{ \varphi (A_{u} + A_{t}) }{\varphi (A_{u})} - 1  \right )}
    \lim_{t \rightarrow x} e^{x_{j} {\partial}_{j}^{t} \left (
  \frac{ \varphi (A_{u} + A_{t}) }{\varphi (A_{t})} e^{A_{u}} - 1
   \right )} f(u) g(t).
\end{equation}
Since the factors in (\ref{d6}) commute, we finally obtain the star
product in the $\varphi$ realization (i.e. in the $\varphi$ ordering)
\begin{equation} \label{d7}
(f \; {\star}_{\varphi} \; g)(x)
  =  \lim_{\substack{u \rightarrow x  \\ t \rightarrow x }}
 e^{x_{j} {\partial}_{j}^{u} \left (
  \frac{ \varphi (A_{u} + A_{t}) }{\varphi (A_{u})} - 1  \right )
     + x_{j} {\partial}_{j}^{t} \left ( 
        \frac{ \varphi (A_{u} + A_{t}) }{\varphi (A_{t})} e^{A_{u}} - 1 \right )}
    f(u) g(t). 
\end{equation}
It should be noted that the coproduct stemming from this result for
the star product completely coincides with the relation (\ref{BCHstar})
and the coproduct (\ref{partialcoproduct}).
 We stress here that the
result (\ref{d7}) is the special case of the more general one
\begin{equation} \label{d88}
(f \; \star \; g)(x)  =   \lim_{\substack{u \rightarrow x  \\ y \rightarrow x }}
 \mu \left ( e^{x_{\alpha} ( \triangle - {\triangle}_{0}) {\partial}_{\alpha} }
    f(u) \otimes g(y) \right ), 
\end{equation}
where $ \; {\triangle}_{0}({\partial}_{\mu}) =
 {\partial}_{\mu} \otimes 1 + 1 \otimes {\partial}_{\mu}, \; $
$ \: \triangle ({\partial}_{\mu}) \; $ is given in (\ref{partialcoproduct})
and $ \; \mu \; $ is the multiplication map in the Hopf algebra,
 namely, $ \; \mu (f(x) \otimes g(x)) = f(x) g(x). $
 The relation (\ref{d88}) 
 can be extended to hold for a general Lie algebra. To end this
 section, note that generally, in view of Eqs.(\ref{BCHordered}) and (\ref{starproduct}), the star product
 $ \; {\star}_{\varphi} \; $ satisfies the relation
   $ \; e^{ik \cdot x}{\star}_{\varphi} e^{iq \cdot x} = e^{i(k \oplus q )\cdot x}, \; $
 with $ \; k \oplus q \; $ defined in (\ref{BCHstar}).


\section{ Towards the scalar field theory on $ \; \kappa$-deformed
  space}

\subsection{$ \; \kappa$-deformed Minkowski space and the
  {\em{natural}} realization}

All results obtained so far are derived for the case of the Euclidean
metric. They can be easily extended to Minkowski space. It is only
 about the change in the definition of the momentum
\begin{equation} \label{impuls}
   P_{0} = i {\partial}_{0},  \;\;\;\;  P_{i} = - i {\partial}_{i},
   \;\;\;\; i = 1,2,...,N-1,
\end{equation}
in order to satisfy
\begin{equation} \label{impuls}
  [P_{\mu}, x_{\nu}] = - i {\eta}_{\mu \nu},  \quad   \mu, \nu = 0,1,..., N-1,
\end{equation}
$ \; {\eta}_{\mu \nu} = diag(-1,+1,+1,+1), \;$
while the algebra of the variables $ \; x_{\mu}, {\partial}_{\nu}  \; $
remains unaltered,
\begin{equation} \label{impuls1}
  [{\partial}_{\mu}, x_{\nu}] = {\delta}_{\mu \nu},  \quad   \mu, \nu = 0,1,..., N-1.
\end{equation}
Accordingly, the commutation relation describing $ \; \kappa$-deformed Minkowski space casts into 
\begin{equation} 
\lbrack\hat{x}_{i},\hat{x}_{j}\rbrack=0, \qquad
\lbrack\hat{x}_{0},\hat{x}_{i}\rbrack=ia\hat{x}_{i},\qquad
i,j=1,2,...,N-1.
\end{equation}
The star product (\ref{d7}) on $ \; \kappa$-deformed
 Minkowski  space also remains unchanged, except that $ \; A_{x} \; $
 now stands for $ \; A_{x} = ia \partial_{0}^{x}. \; $ It is important
 to stress that the deformation parameter $ \; a \; $ appearing in 
 the above commutation relation is not the same as the one
  we were dealing with when we considered the Euclidean space. They are
 related as $ \; a_{n}=ia_{0}, \; $ 
 where $ \; a_{n} \; $ is the old
 deformation parameter and $ \; a_{0} \; $ is the new one.
For simplicity, we remove the index 0 of it, but in all subsequent
 consideration one should keep in
 mind that it is the time component of the $ \; n$-vector in Minkowski space
and that it is real. The interchange $ \; a_{n}=ia_{0}, \; $ together with the whole set of
 relations $ \; x_{n}=ix_{0}, \; D_{n}=-iD_{0}, \; M_{in}=-iM_{i0} \; $
provides the consistent transition from Euclidean to Minkowski space.

Among all mappings that realize $ \; \kappa$-deformed Minkowski
space, there is one of special significance because, unlike the
general mapping (\ref{mapping}), it is covariant. It has some
interesting properties and is introduced in a natural way 
by identifying the Dirac covariant derivative with the ordinary
derivative $ \; D_{\mu} \equiv {\partial}_{\mu}. \; $
This special kind of realization has already been introduced at the end of section 2.2 through the
mapping (\ref{natrellie}) to which we have assigned the term {\em natural} realization.
Since we are interested in this kind of realization with respect to
the deformed Heisenberg algebra (\ref{com11}), according to (\ref{com11}), we have
\begin{equation}
\lbrack
  {\partial}_{\mu},\hat{x}_{\nu}^{Nat} \rbrack = \delta_{\mu \nu}(-ia
  {\partial}_{0} + \sqrt{1+a^2 {\partial}^{2}}) + i a \delta_{\mu 0} {\partial}_{\nu},
\end{equation}
yielding the following realization of the noncommuting coordinates:
\begin{equation} \label{natrel}
  \hat{x}_{\mu}^{Nat}  = x_{\mu}(-ia
  {\partial}_{0} + \sqrt{1+a^2 {\partial}^{2}}) + i a x_{0} {\partial}_{\mu}.
\end{equation}

Thus, in the following (\ref{natrel}) will be referred to as a {\em natural} realization
and it will be understood that it is natural with respect to the deformed Heisenberg 
algebra (\ref{com11}). We recall that (\ref{com11}) is fixed since it is a
part of the extended Lorentz algebra (see section 2.1).
 Note that the mapping (\ref{natrel}) is not incorporated in the set of realizations (\ref{mapping}).
The properties of the {\em natural} realization are expressed in a
 most convenient way if the Dirac covariant derivative (\ref{com7}) is associated
 with the momentum $ \; P_{\mu}^{\varphi} \; $ on $ \; \kappa$-deformed Minkowski  space,
\begin{equation} \label{deformedimpuls}
\begin{array}{c}
P_{i}^{\varphi} = -iD_{i}= - i\partial_{i} \frac{e^{-A}}{\varphi (A)}
   = k_{i} \frac{e^{-ak_{0}}}{\varphi (ak_{0})} ,\\
\\
P_{0}^{\varphi} = iD_{0}= i\partial_{0} \frac{\sinh A}{A} + a \Delta \frac{e^{-A}}{2\varphi^2} =
     \frac{1}{a} \sinh (ak_{0}) - a {\vec{k}}^{2} \frac{e^{-ak_{0}}}{2\varphi^2},
\end{array}
\end{equation}
where $ \; k_{0} = i \partial_{0}, \;\; k_{i} = -i \partial_{i}. \; $
This association can be regarded as a map between the
algebra of fields on noncommutative $ \; \kappa$-Minkowski
space and the algebra of ordinary fields on Minkowski spacetime
equipped with a star product \cite{22},\cite{23}. 
 The special case when $\; \varphi = 1, \;$ which is the case of right
 ordering, was considered in \cite{26}.
 Since the map (\ref{deformedimpuls}) is fully
specified by its action on plane waves, it can be looked on as if it
mapped ordinary Minkowski plane waves $ \; e^{ik_{\alpha} x^{\alpha}} \; $
 to plane waves 
$ \; :e^{iP_{\alpha}(k) {\hat{x}}^{\alpha}}:_{\varphi} \; $
on $ \; \kappa$-deformed Minkowski space, $ \; P: \; k \mapsto P(k), \; $
with $ \; P(k) \; $ given by (\ref{deformedimpuls}).

The {\em natural} realization has a property
\begin{equation} \label{naturalformula}
 e^{ik^{\alpha} {\hat{x}}_{\alpha}^{Nat}} |0 \rangle =
 e^{iP_{\alpha}^{S} (k) x^{\alpha}}, 
\end{equation}
where $ \; P_{\alpha}^{S} \; $ is the momentum (\ref{deformedimpuls})
evaluated for $ \; \varphi = {\varphi}_{S}, \; $ i.e. corresponding to
symmetric ordering. From this and the relations (\ref{deformedimpuls})
  it is easy to calculate the following expression:
\begin{equation} \label{naturalformula1}
  :e^{i k^{\alpha} {\hat{x}}_{\alpha}^{Nat}}:_{\varphi} |0 \rangle =
 e^{i( -k_{0} {\hat{x}}_{0}^{Nat} + k_{i} {\hat{x}}_{i}^{Nat} \frac{{\varphi}_{S}(ak_{0})}
 {\varphi(ak_{0})})} |0 \rangle \equiv
 e^{i K_{\varphi}^{\alpha}(k) {\hat{x}}_{\alpha}^{Nat}} |0 \rangle 
 = e^{i P^{\varphi, \alpha}(k) {x}_{\alpha}},
\end{equation}
with $ \; K_{\varphi}(k) = (k_{0}, k_{i} \frac{{\varphi}_{S}(ak_{0})} {\varphi(ak_{0})} ) \; $
and $ \; P_{\alpha}^{\varphi} \; $ given in
(\ref{deformedimpuls}). The function $ \; K_{\varphi}(k) \; $ is an
example of the map introduced in (\ref{Kmapping}).


\subsection{Free scalar field theory on $ \; \kappa$-deformed
  Minkowski space}

The {\em{natural}} realization appears as a particularly convenient
one for constructing the field theory on noncommutative space. The
reason for this is that it enables one to recast the action belonging
to an algebra of functions whose product is a noncommutative star
product into the action expressed entirely in terms of usual pointwise
multiplication between two functions. 
We shall see that although the construction that is to be made
requires one particular noncommutative star product on an algebra of
functions, namely the one which is related to the {\em{natural}}
realization, the realization of the NC space alone is
not important at all. In other words, the realizations of 4-momenta
and coordinates on NC space are not important in constructing
the noncommutative field theory action. The final result, obtained
from  the original action by virtue of the properties of
{\em{natural}} star product $ \; {\star}_{N}, \; $ will be cast into the form of the nonlocal,
relativistically invariant field theory whose action is expressed in
terms of usual pointwise multiplication of fields and has
particularly simple form.
This result is the same regardless of the particular realization used
to represent 4-momenta and coordinates on NC space. 

 So far we have seen that each realization of NC space was related to
 some specific type of the ordering prescription. The same situation
 as well
  appears to be true in the case of the {\em{natural}} realization.
As was the case with the realizations
 (\ref{mapping}), where the corresponding orderings are expressed in
 terms of the exponentials of the type
\begin{equation} \label{orderingminkowski}
 :e^{ik^{\alpha} {\hat{x}}_{\alpha}}:_{\varphi}  =
 e^{-ik_{0} {\hat{x}}_{0} + ik_{i} {\hat{x}}_{i} \frac{{\varphi}_{S}(ak_{0})}
 {\varphi(ak_{0})}}, 
\end{equation} 
(The relation (\ref{orderingminkowski}) is the same as
(\ref{ordering}), the difference being only in the signature adjusted to hold for
the Minkowski space) the similar connection can be found for the
{\em{natural}} realization either. In fact, it can immediately be
deduced from Eq.(\ref{naturalformula}) by rearranging it to obtain
\begin{equation} \label{naturalordering}
   e^{i {(P^{S})}^{-1}(k)\cdot {\hat{x}}^{Nat} } |0 \rangle =
   e^{i k \cdot x}.
\end{equation}
In order to specify the function $ \; {(P^{S})}^{-1}, \; $
appearing in (\ref{naturalordering}), we once again write the relations (\ref{deformedimpuls}),
\begin{equation} \label{deformedimpuls1}
\begin{array}{c}
P_{i}^{\varphi} =
    k_{i} \frac{e^{-ak_{0}}}{\varphi(ak_{0})} ,\\
\\
P_{0}^{\varphi} = 
     \frac{1}{a} \sinh (ak_{0}) - a {\vec{k}}^{2}
    \frac{e^{-ak_{0}}}{2\varphi^2 (ak_{0})}.
\end{array}
\end{equation}
By inverting them for the case when $ \; \varphi = {\varphi}_{S},\; $ we obtain
the required quantity $ \; {(P^{S})}^{-1} \; $  in components
\begin{equation} \label{inversedeformedimpuls1}
\begin{array}{c}
 {(P^{S})}^{-1}_{0}(P) = 
    \frac{1}{a} \ln \frac{a P_{0} + \sqrt{1 - a^2 P^{2} } }
           {1 - a^2  {\vec{P}}^{2}},\\
\\
  {(P^{S})}^{-1}_{i}(P) = 
  P_{i} {\varphi}_{S} \left ( \ln \frac{a P_{0} + \sqrt{1 - a^2 P^{2} } }
           {1 - a^2  {\vec{P}}^{2}} \right )
             \frac{a P_{0} + \sqrt{1 - a^2 P^{2} } }
           {1 - a^2  {\vec{P}}^{2}}.
\end{array}
\end{equation}
The expression (\ref{naturalordering}), together with
 (\ref{inversedeformedimpuls1}), defines the ordering
 prescription corresponding to the {\em natural}
 realization.

We now turn to the notion of
 the star product corresponding to the {\em{natural}}
 realization. This star product, for which we adopt the designation
 $ \; {\star}_{N}, \; $ is characterized by the property
\begin{equation} \label{keystarproductformula}
  e^{i P^{\varphi}(k)\cdot x} {\star}_{N} e^{i P^{\varphi}(q)\cdot x} =
   e^{i P^{\varphi}(k \oplus q)\cdot x},
\end{equation}
where 
\begin{equation} \label{coproductformula}
 k \oplus q = \left ( k_{0} + q_{0}, \;\; k_{i} \frac{\varphi(ak_{0} + aq_{0}) }{\varphi(ak_{0})}
   + q_{i} \frac{\varphi(ak_{0} + aq_{0}) }{\varphi(aq_{0})}
   e^{ak_{0}} \right )
\end{equation}
is the generalized rule for the addition of momenta
and $ \; P^{\varphi} \; $ is defined in (\ref{deformedimpuls}).
It is understood that the quantity $ \; k \oplus q \; $ is taken in
the $ \; \varphi \; $ realization, although it is not specificaly indicated.
The relation (\ref{keystarproductformula}) will have a crucial role in
our construction of noncommutative field theory. From the
definition of the inverse 4-momentum $ \; k \oplus k^{-1} = (0,
\vec{0}), \; $ if $ \; k  = (k_{0}, \vec{k}), \; $ then we have
\begin{equation} \label{inverseformula}
 k^{-1} = \left ( - k_{0}, \;\; - k_{i} \frac{\varphi(- ak_{0}) }{\varphi(ak_{0})} 
   e^{- ak_{0}} \right ),
\end{equation}
\begin{equation} \label{inversecoproductformula}
 k^{-1} \oplus q = \left (- k_{0} + q_{0}, \;\; 
  \left ( - \frac{ k_{i}}{\varphi(ak_{0})} + \frac{
  q_{i}}{\varphi(aq_{0})} \right )
    \varphi(- ak_{0} + aq_{0})  e^{- ak_{0}} \right ).
\end{equation}
In the calculations that follow, we adopt the Hermitean conjugation property 
\begin{equation} \label{conjugation}
  { \left ( e^{i P^{\varphi}(k) \cdot x} \right )}^{\dagger } =
   e^{i P^{\varphi}(k^{-1}) \cdot x},
\end{equation}

Let us now derive the relation (\ref{keystarproductformula}) which encloses the property of {\em natural}
 star product important for our construction.
Utilizing Eqs.(\ref{BCHordered}) and (\ref{naturalformula1}), we have 
$$ 
  e^{i P^{\varphi}(k)\cdot x} {\star}_{N} e^{i P^{\varphi}(q) \cdot x} 
$$
$$
 = :e^{i k \cdot {\hat{x}}^{Nat}}:_{\varphi}
     :e^{i q \cdot {\hat{x}}^{Nat}}:_{\varphi} |0 \rangle
$$
$$
  = :e^{i (k \oplus q) \cdot {\hat{x}}^{Nat}}:_{\varphi} |0 \rangle 
$$
\begin{equation} \label{keystarderivation}
   = e^{i P^{\varphi}(k \oplus q) \cdot x},
\end{equation}
 thus completing the proof of (\ref{keystarproductformula}).
In the second line of the derivation (\ref{keystarderivation}), we have
applied Eq.(\ref{naturalformula1}), while in the
third line we have used the property (\ref{BCHordered}) satisfied
by the generalized addition of momenta
 (adjusted for the Minkowski signature). Finaly, in the
last line, Eq.(\ref{naturalformula1}) is used again, but in a
reverse direction.

The result (\ref{keystarproductformula}) can be further used to derive
the identity
\begin{equation} \label{keyformula}
  \int d^{n}x   { \phi}^{\dagger}(x) {\star}_{N} \psi (x) =
     \int d^{n}x   { \phi}^{\ast}(x)   
   \sqrt{1 + a^2 {\partial}_{\mu} {\partial}^{\mu} } \psi (x),
\end{equation}
which provides the transition from the integration of the star product of two
fields to the integration of the ordinary product of fields. To prove it,
it is sufficient to show that it holds for plane waves
 $ \; \phi (x) = e^{iP^{\varphi}(k) \cdot x}, \; \psi (x) = e^{iP^{\varphi}(q) \cdot x}. \; $  
Thus, from (\ref{keystarproductformula}) and (\ref{conjugation})  we have
\begin{equation} \label{keyformulaa1}
  \int d^{n}x  { \left ( e^{i P^{\varphi}(k) \cdot x} \right )}^{\dagger }
 {\star}_{N} e^{i P^{\varphi}(q) \cdot x} = \int d^{n}x e^{i
 P^{\varphi}(k^{-1} \oplus q) \cdot x} = {(2 \pi)}^{n} \;  
 \delta \left ( P_{0}^{\varphi} (k^{-1} \oplus q) \right )
            {\delta}^{(n-1)} \left ( {\vec{P}}^{\varphi} (k^{-1} \oplus q) \right ).
\end{equation}
In order to perform the calculations on the right-hand side of
(\ref{keyformulaa1}), it is convenient to introduce the quantities
\begin{equation} \label{pcetiri}
\begin{array}{c}
 {\mathcal{P}}_{n}^{\varphi} (k) = 
    - \frac{1}{a} \cosh (ak_{0}) + a {\vec{k}}^{2}
    \frac{e^{-ak_{0}}}{2 \varphi^2 (ak_{0})},\\
\\
 P_{\pm}^{\varphi} = {\mathcal{P}}_{n}^{\varphi} \pm P_{0}^{\varphi}
\end{array}
\end{equation}
(The first quantity in (\ref{pcetiri}) should not be confused with the
 Euclidean counterpart $ \; P_{n}= i P_{0}  \; $ of $ \; P_{0}. $ )
It is now straightforward to obtain the following expressions:
\begin{equation} \label{pcetiri2}
\begin{array}{c}
 P_{i}^{\varphi} (k^{-1} \oplus q) = 
    \left ( \frac{q_{i}}{\varphi(aq_{0})} -
    \frac{k_{i}}{\varphi(ak_{0})} \right ) e^{-aq_{0}},\\
\\
 P_{0}^{\varphi} (k^{-1} \oplus q) =
   \frac{1}{a} \sinh (aq_{0} - ak_{0}) - \frac{a}{2}
   { \left ( \frac{\vec{q}}{\varphi(aq_{0})} -
    \frac{\vec{k}}{\varphi(ak_{0})} \right ) }^{2} e^{-ak_{0} - aq_{0}},\\
 \\ 
 P_{+}^{\varphi}(k) = - \frac{1}{a} e^{-ak_{0}}.
\end{array}
\end{equation}
When they are combined with (\ref{deformedimpuls1}), we get
\begin{equation} \label{pcetiri3}
\begin{array}{c}
 P_{i}^{\varphi} (k^{-1} \oplus q) = 
    \frac{P_{+}^{\varphi} (k) P_{i}^{\varphi} (q) - P_{i}^{\varphi}
    (k) P_{+}^{\varphi} (q)}{P_{+}^{\varphi} (k)},\\
\\
 P_{0}^{\varphi} (k^{-1} \oplus q) =
   -a \bigg[ P_{+}^{\varphi} (k) P_{0}^{\varphi} (q) -
    P_{0}^{\varphi} (k) P_{+}^{\varphi} (q) - P_{i}^{\varphi}(k)
        P_{i}^{\varphi} (k^{-1} \oplus q) \bigg].
\end{array}
\end{equation}
Since the expression (\ref{keyformulaa1}) vanishes for
 $ \; \vec{P}^{\varphi} (k^{-1} \oplus q) \neq 0, \; $ the last
term of $ \; P_{0}^{\varphi} (k^{-1} \oplus q) \; $ in (\ref{pcetiri3}) drops out
and we are left with   
\begin{equation} \label{pcetiri4}
 P_{+}^{\varphi} (k) P_{0}^{\varphi} (q) -
    P_{0}^{\varphi} (k) P_{+}^{\varphi} (q) =
  \frac{1}{2 a^2 P_{+}^{\varphi} (k) P_{+}^{\varphi} (q)}
  \bigg[ { P_{+}^{\varphi} (q) }^{2} - { P_{+}^{\varphi} (k) }^{2}
 + a^2 {( P_{+}^{\varphi} (k) \vec{P}^{\varphi}(q) )}^{2} 
 -  a^2 {( P_{+}^{\varphi} (q) \vec{P}^{\varphi}(k) )}^{2} \bigg],
\end{equation} 
where use has been made of the identity
$ \; 2P_{0}^{\varphi} = P_{+}^{\varphi} -
 \frac{1}{a^2 P_{+}^{\varphi}} + \frac{{\vec{P}_{\varphi}}^{2}}{P_{+}^{\varphi}}. \; $
From the same argument as before, the last two terms in (\ref{pcetiri4}) cancel each other,
leaving us with
$$
 {(2 \pi)}^{n} {\delta}^{(n)} \left ( P^{\varphi} (k^{-1} \oplus q) \right ) =
  {(2 \pi)}^{n} \delta \left ( - a
 \frac{{ P_{+}^{\varphi} (q) }^{2} - { P_{+}^{\varphi} (k) }^{2}}
   {2 a^2 P_{+}^{\varphi} (k) P_{+}^{\varphi} (q)} \right )
 {\delta}^{(n-1)} \left ( {\vec{P}}^{\varphi} (q) -
 \frac{P_{+}^{\varphi} (q)}{P_{+}^{\varphi} (k)} {\vec{P}}^{\varphi} (k) \right )
$$ 
$$
 = {(2 \pi)}^{n} 
    \frac{ \delta \left (P_{+}^{\varphi}(k) - P_{+}^{\varphi}(q) \right )}
      {\Big| - \frac{{ P_{+}^{\varphi} (k) }^{2} + { P_{+}^{\varphi} (q) }^{2}}
   {2 a^2 P_{+}^{\varphi} (q) {P_{+}^{\varphi} (k)}^{2}} (-a)
 \Big|_{P_{+}^{\varphi}(k) = P_{+}^{\varphi}(q)}} 
 {\delta}^{(n-1)} \left ( {\vec{P}}^{\varphi} (q) -
 \frac{P_{+}^{\varphi} (q)}{P_{+}^{\varphi} (k)} {\vec{P}}^{\varphi} (k) \right )
$$ 
\begin{equation} \label{pcetiri5}
   = {(2 \pi)}^{n} a P_{+}^{\varphi} (q) \delta ( P_{+}^{\varphi}
      (k) - P_{+}^{\varphi} (q) )
      {\delta}^{(n-1)} \left ( {\vec{P}}^{\varphi} (q) -
  {\vec{P}}^{\varphi} (k) \right ).
\end{equation}
Here, in calculating the above expression, one zero of
 the function in the argument has been rejected due to the requirement
that $ \; P_{+}^{\varphi} < 0. \; $ This condition is obvious from the
 form of $ \; P_{+}^{\varphi} \; $ in (\ref{pcetiri2}).
Since the final goal is to express the result in terms of $ \; P_{0}^{\varphi} \; $
and $ \; \vec{P}^{\varphi}, \; $ we perform the following computation
$$
 \delta \left ( P_{0}^{\varphi}(q) - P_{0}^{\varphi}(k) \right ) =
    \delta \left ( \frac{1}{2}P_{+}^{\varphi}(q) - \frac{1}{2 a^2
 P_{+}^{\varphi} (q)} + \frac{{{\vec{P}}^{\varphi}(q)}^{2}}{2 P_{+}^{\varphi}(q)}
  - \frac{1}{2}P_{+}^{\varphi}(k) + \frac{1}{2 a^2 P_{+}^{\varphi}(k)} -
  \frac{{{\vec{P}}^{\varphi}(k)}^{2}}{2 P_{+}^{\varphi}(k)} \right ) 
$$ 
\begin{equation} \label{pcetiri6}
  = \frac{ \delta \left (P_{+}^{\varphi}(k) - P_{+}^{\varphi}(q) \right )}
      {\Big| - \frac{1}{2} - \frac{1}{2 a^2 {P_{+}^{\varphi}(k)}^{2}}
  + \frac{{{\vec{P}}^{\varphi}(k)}^{2}}{2 {P_{+}^{\varphi}(k)}^{2}}
 \Big|_{P_{+}^{\varphi}(k) = P_{+}^{\varphi}(q)}} =
 \frac{|P_{+}^{\varphi}(q)|}{|{\mathcal{P}}_{n}^{\varphi}(q)|} \delta \left
 (P_{+}^{\varphi}(k) - P_{+}^{\varphi}(q) \right ),
\end{equation}
where use has been made of the relation $ \; {\mathcal{P}}_{n}^{\varphi}(k) =
\frac{1}{2} P_{+}^{\varphi}(k) + \frac{1}{2 a^2 P_{+}^{\varphi}(k)} -
 \frac{{{\vec{P}}^{\varphi}(k)}^{2}}{2 P_{+}^{\varphi}(k)}. \; $
Inserting (\ref{pcetiri6}) in (\ref{pcetiri5}), we get
\begin{equation} \label{pcetiri7}
 {(2 \pi)}^{n} {\delta}^{(n)} \left ( P^{\varphi} (k^{-1} \oplus q) \right ) =
  {(2 \pi)}^{n}  a |{\mathcal{P}}_{n}^{\varphi} (q)|
 \delta \left ( P_{0}^{\varphi}(q) - P_{0}^{\varphi}(k) \right )
 {\delta}^{(n-1)} \left ( {\vec{P}}^{\varphi} (q) -
  {\vec{P}}^{\varphi} (k) \right )
\end{equation}
and it can be recognized as  
\begin{equation} \label{pcetiri8}
  {(2 \pi)}^{n}  a |{\mathcal{P}}_{n}^{\varphi} (q)|
 \delta \left ( P_{0}^{\varphi}(q) - P_{0}^{\varphi}(k) \right )
 {\delta}^{(n-1)} \left ( {\vec{P}}^{\varphi} (q) -
  {\vec{P}}^{\varphi} (k) \right ) = a \int  
   d^{n}x  { \left ( e^{i P^{\varphi}(k) \cdot x} \right )}^{\ast }
 |{\mathcal{P}}_{n}^{\varphi} (q)| e^{i P^{\varphi}(q) \cdot x}.
\end{equation}
Finally, the result we have obtained is
\begin{equation} \label{pcetiri8}
 \int d^{n}x  { \left ( e^{i P^{\varphi}(k) \cdot x} \right )}^{\dagger }
 {\star}_{N} e^{i P^{\varphi}(q) \cdot x}  
 =  \int  
   d^{n}x  { \left ( e^{i P^{\varphi}(k) \cdot x} \right )}^{\ast }
 \sqrt{1 + a^2 {\partial}_{\mu} {\partial}^{\mu}} e^{i P^{\varphi}(q) \cdot x}.
\end{equation}
In the last step, we have utilized the identity 
$ \; a | {\mathcal{P}}_{n}^{\varphi}| =
 \sqrt{1 - a^2 {P^{\varphi}}^{2} } = \sqrt{1 + a^2 D^{\mu}D_{\mu} } \; $
and the fact that we are working in the {\em natural} realization where 
$ \; D_{\mu} \equiv {\partial}_{\mu}. \; $
This proves the relation (\ref{keyformula}) which is a generalization
of the integral identity found in \cite{26}.
Here it was proved by starting from the general ordering prescription characterized by the
 function $\; \varphi, \;$ in contrast to Ref.\cite{26} where the same
 relation is obtained by starting from some particular ordering prescription,
 namely, the right one.

We can now look upon the action for a free massive scalar field on 
noncommutative $ \; \kappa$-deformed Minkowski space 
\begin{equation} \label{action}
 S = \int d^{4}x \bigg[ \frac{1}{2} {({\partial}_{\mu} \phi)}^{\dagger} {\star}_{N}
         ({\partial}^{\mu} \phi) (x) + \frac{m^2}{2} {\phi}^{\dagger}
         {\star}_{N} \phi (x) \bigg].
\end{equation}
The construction of this action is based on the star product
corresponding to the {\em natural} realization. At this place we point
out that the action (\ref{action}) is $\; \kappa$-Poincare
invariant. Namely, in the {\em natural} realization, where $\; D_{\mu}
= {\partial}_{\mu}, \;$ the Dirac derivative  $\; D_{\mu} \;$ is a
vector under Lorentz algebra and the coproduct $\; {\triangle}_{N} (
D_{\mu}) \;$ transforms in a covariant way  \cite{23}. This coproduct
(for the Euclidean case) is given by 
\begin{equation} \label{coproductndemi}
  {\triangle}_{N} (D_{\mu}) = D_{\mu} \otimes Z^{-1} + 1 \otimes
  D_{\mu} + i a_{\mu} D_{\lambda} Z \otimes D_{\lambda} - \frac{i
  a_{\mu}}{2} \Box Z \otimes (i a_{\lambda} D_{\lambda}), 
\end{equation}
where
\begin{equation} 
 Z^{-1} = - i a_{\lambda} D_{\lambda} + \sqrt{1 + a^2 D_{\lambda}D_{\lambda}}.
\end{equation}
The last equation represents an invariant shift operator \cite{23}.
From expressions (see Eq.(\ref{d88})) 
$$
(f  {\star}_N  g)
(u) = \mu \Big( e^{u_{\alpha}({\triangle}_N -
\triangle_0) {\partial}_{\alpha}} (f\otimes
g) (x) \Big) \Big|_{x = u}.
$$
and
$$
 e^{ik \cdot x} {\star}_N e^{iq \cdot x} = e^{i {\triangle}_N (k,q) \cdot x}
$$
we conclude that $\; {\star}_N \;$ product in {\em natural}
realization (where $\; D_{\mu} = {\partial}_{\mu} $) is
noncommutative, nonlocal, but $\; \kappa$-Poincare
invariant. Hence, the star product $\; {\star}_N \;$ of two scalar
fields is scalar with respect to $\; \kappa$-Poincare
symmetry. Similarly, from the same argument, the term 
$\; D_{\mu} \phi {\star}_N  {D}^{\mu} \phi \; $ is also invariant
under the action of $\; \kappa$-Poincare symmetry. As a consequence of
these consideration, we conclude that the action (\ref{action}) is
invariant under the Poincare algebra (see also Eq.(\ref{keyformula})) and $\; \kappa$-Poincare
symmetry.

 A few
comments are in order regarding expression (\ref{action}) . Considering the action (\ref{action}) or the
expression (\ref{pcetiri8}) obtained by spliting (\ref{action}) to
Fourier components, we see that the construction of the action can be basicaly
taken along two different routes. The first one keeps the realizations of
4-momenta appearing in (\ref{pcetiri8}) fixed, while
simultaneously changing the ordering prescription which the star
product should be associated to. The second route is based on
choosing the convenient ordering prescription with its accompanying
star product and keeping it fixed, while simultaneously changing the
realizations of 4-momenta appearing in the exponentials of
(\ref{pcetiri8}) and used to define $\; \kappa$-Minkowski space.
 Here we adopt the second approach,
where we choose the star product corresponding to {\em natural}
realization and keep it fixed.
 This realization of the
deformed space has special properties which enable us to cast the
theory described by the action (\ref{action}) in a form which
 is easier to handle and which is closer in form to
theories we know how to deal with. 
With the help of the identity (\ref{keyformula}) relating the
integration of the star
product of fields to the integration of the standard pointwise product of fields, 
the action (\ref{action}) can be cast in the form
\begin{equation} \label{action1}
 S = \int d^{4}x \bigg[ \frac{1}{2} {({\partial}_{\mu} \phi)}^{\ast}
         \sqrt{1 + a^2 {\partial}_{\mu} {\partial}^{\mu} }
         ({\partial}^{\mu} \phi) (x) + \frac{m^2}{2} {\phi}^{\ast}
         \sqrt{1 + a^2 {\partial}_{\mu} {\partial}^{\mu}} \phi (x) \bigg].
\end{equation}
 This shows that the free scalar field theory on $ \; \kappa$-deformed
 Minkowski space is equivalent to a nonlocal, relativistically
 invariant free scalar field theory on Minkowski spacetime. Also, we
 see that
 once the ordering prescription together with its noncommutative star
 product is chosen and fixed, the resulting action (\ref{action1}) is
 the same, independent of the ``gauge'' function $\; \varphi, \; $
 regardless of the realization of 4-momenta in the corresponding
 NC space. Although the theory, through the square root, has higher
 derivatives, thus being nonlocal, it poses no significant physical effects as long as  
 the interaction is not included. Particularly, we expect that there
 will be no new poles leading to the problems with unitarity. However,
 at the quantum level one would come up to the necessity for the
 change in statistics, causing the situation to be more complex.

 It is worth mentioning that the $\; \kappa$-type of the space
 deformation can also be approached on the basis of an alternative general treatment \cite{40}
 of twists with the star products.
 This includes the investigation of the particular type of the
  map, deforming the commutative subalgebra into a noncommutative one.
The deformation map of this kind has already been applied \cite{40} to
 yield the $\; n$-dimensional Moyal plane, as well as the Manin plane.

   At the end, let us make some comments. In the present paper, we
   have started from the $\; \kappa$-deformed algebra (\ref{com1}) and
   considered its extension to the extended Lorentz algebra where we have included
   the generators of rotations and the covariant Dirac derivative $\; D_{\mu}.$
  The requirement that the Lorentz algebra is undeformed caused the
   deformed Heisenberg algebra (\ref{com11}) to be uniquelly determined.
 For this deformed Heisenberg algebra there exists a unique 
{\em natural} realization, specified by the requirement $\; D_{\mu}
   \equiv {\partial}_{\mu}, \;$ and given by the relations
 (\ref{natrellie}), i.e. (\ref{natrel}). There is also a unique
   ordering prescription corresponding to this realization. It was refered to as
   a natural ordering and it was specified by (\ref{naturalordering}).
 We were also found the coproduct and the star product $\; {\star}_{N} \;$ corresponding
   to the {\em natural} realization. The star product $\; {\star}_{N} \;$ 
 has been shown to obey the relation (\ref{keystarproductformula}) which implied
 the following property of $\; {\star}_{N} \;$ to be satisfied
$$
  \int d^{n}x   { \phi}_{\varphi}^{\dagger}(x)  {\star}_{N}
  {\psi}_{\varphi}  (x) =
 \int d^{n}k \; {\tilde{\phi}}^{\ast}(k) \int d^{n}q \; \tilde{\psi}(q) \int d^{n}x 
 { \left ( e^{i P^{\varphi}(k) \cdot x} \right )}^{\dagger }
 {\star}_{N} e^{i P^{\varphi}(q) \cdot x} 
$$
\begin{equation} \label{keyformulageneral}
  = \int d^{n}k \; {\tilde{\phi}}^{\ast}(k) \int d^{n}q \; \tilde{\psi}(q)
 \int d^{n}x :e^{i k^{-1} \cdot {\hat{x}}^{Nat}}:_{\varphi} 
     :e^{i q \cdot {\hat{x}}^{Nat}}:_{\varphi} |0 \rangle
     = \int d^{n}x  \; { \phi}_{\varphi}^{\ast}(x) \;  
   \chi (\partial) \; {\psi}_{\varphi} (x),
\end{equation}
where, according to (\ref{keyformula}), the function $\; \chi \;$ is
$\; \chi (\partial) = \sqrt{1 + a^2 {\partial}^{2} }. \;$ In writing the
Fourier expansions of the fields $\; \phi \;$ and $\; \psi, \;$ a
particular realization $\; \varphi \;$ of NC space was used, but the
final result, i.e. the function $\; \chi \;$ does not depend on
the realization $\; \varphi.$ In the steps undertaken in
(\ref{keyformulageneral}) (which are, in fact, concise
summary of the calculations done earlier in the paper, see
particularly Eqs.(\ref{naturalformula1}),(\ref{keystarderivation}) and
the consideration after that), we have utilised
the following field expansion
$$
 {\phi}_{\varphi} (x) = \int d^{n}x \tilde{\phi}(k) e^{i P^{\varphi}(k)
 \cdot x}. 
$$
It is consistent, through the relation (\ref{naturalformula1}),
 with the Fourier expansion (\ref{Fouriertransform})
$$
 {\phi}_{\varphi} (\hat{x}) = \int d^{n}x \tilde{\phi}(k) {:e^{i k
 \cdot \hat{x}} :}_{\varphi}
$$
and the relation $\; {\phi}_{\varphi} ({\hat{x}}^{Nat}) |0 \rangle =
 {\phi}_{\varphi}^{Nat} (x) \equiv {\phi}_{\varphi} (x).$

   It is important to emphasize that we could equally well start our consideration from
   an arbitrary deformed Heisenberg algebra
\begin{equation} \label{generalheisenberg}
  [D_{\mu}, {\hat{x}}_{\nu} ] = {\varphi}_{\mu \nu} (D),
\end{equation}
where $\; {\varphi}_{\mu \nu} \;$ is some function compatible with the
commutation relations (1),
and take the same steps as those performed for the algebra
(\ref{com11}). We would also find the unique {\em natural} realization
(characterized by $ \; D_{\mu} \equiv {\partial}_{\mu}  \; $ and the mapping
$ \; {\hat{x}}_{\mu}^{Nat} = x_{\alpha} {\varphi}_{\alpha
  \mu}(\partial) \; $ which has to satisfy the commutation relations (1))
and {\em natural} ordering corresponding to the
algebra (\ref{generalheisenberg}). There would also be a coproduct and a star
product $\; {\star}_{N} \;$ which could be associated with (\ref{generalheisenberg}).
One would find that the specific property of the star product $\; {\star}_{N}, \;$ 
corresponding to (\ref{generalheisenberg}), is also enclosed in the
identity (\ref{keyformulageneral}), except for the fact that the
function $\; \chi \;$ now has a form appropriate to the case in
consideration. 

The adventage of our approach is in a simple construction of  $\;
\kappa$-Poincare action using the notion of {\em natural}
realization.
This route of consideration has an advantage that
it provides a direct way for constructing any $\; \kappa$-Poincare
invariant action.
 The Dirac derivative $ \; D_{\mu}, \; $ Eq.(\ref{com7}),
transforms as a vector and the right hand sides of the commutation relations
(\ref{com4}),(\ref{com5}) and (\ref{com11}) have covariant form. Hence, the
coproducts  $\; \triangle (D_{\mu}) \;$ and $\; \triangle (M_{\mu
  \nu}) \;$ have also covariant forms under $\; \kappa$-Poincare
symmetry. Thus we have that starting with generators of undeformed Poincare algebra, the
right hand sides of (\ref{com4}),(\ref{com5}) and (\ref{com11}) and the
corresponding coproducts  $\; \triangle (D_{\mu}) \;$ and $\; \triangle (M_{\mu
  \nu}) \;$ are covariant under the twisted Poincare symmetry.

In papers \cite{19},\cite{24},\cite{Grosse:2005iz} the symmetric
realization corresponding to symmetric Weyl ordering was used, but
therein $\; \triangle (D_{\mu}) \;$ and $\; \triangle (M_{\mu \nu})
\;$ are given by Eqs.(\ref{trt11}),(\ref{com7}) with $\; \varphi (A) =
\frac{A}{e^A -1}, \;$ whereas in our approach, which uses the notion
of {\em natural} realization, $\; D_{\mu} \;$ and $\; M_{\mu \nu} \;$
have particularly simple form \cite{23}, namely $\; D_{\mu} =
{\partial}_{\mu} \;$ and $\; M_{\mu \nu} = {x}_{\mu} {\partial}_{\nu}
- {x}_{\nu} {\partial}_{\mu} .$ Hence, both realizations are covariant
and equivalent. In analogy with the gauge theory they correspond to
different covariant gauge conditions. Finaly, in contrast to
\cite{19},\cite{Grosse:2005iz}, where the deformed Klein-Gordon operator has
the form $\; \Box + m^2 = \frac{2}{a^2} (1 - \sqrt{1 - a^2 D^2} ) +
m^2, \;$
we take the simplest possibility for the Klein-Gordon
operator by choosing  $\; \Box + m^2 = D^2 + m^2.$ 
 The physical consequences of this approach are
still under investigation and will be reported in near future.


\section{Conclusion}

We have constructed $ \; \kappa$-deformed noncommutative
space-time of the Lie algebra type with undeformed extended Lorentz algebra
and deformed coalgebra by realizing noncommutative coordinates,
generalized derivatives and generalized rotation generators in terms
of the commutative coordinates and their derivatives.
 The coproducts of the generalized rotation
generators and of the Dirac derivatives, respectively, have been
identified. A unique correspondence between the particular
realization of $ \; \kappa$-deformed space-time and the corresponding ordering
prescription has been found. For fixed deformed Heisenberg algebra (\ref{com11})
 all these realizations of $ \; \kappa$-
deformed Minkowski space are of equal importance, none of the realizations
is preferable to the others, implying that physical results do
not depend on the particular realization chosen. Accordingly, the choice
of the ordering prescription or the choice of a particular realization
is analogous to the situation in gauge theory where a particular gauge is chosen. We have
found the expression for the star product on $ \; \kappa$-Minkowski
space corresponding to an arbitrary ``gauge'' function $\; \varphi \;$ 
defining the realization. This result is related to the important
relation between a coproduct and a star product in terms of an
exponential map.

 Although all realizations are on equal footing, for
constructing a field theory on $ \; \kappa$-deformed Minkowski space,
we have used a particular realization, the {\em{natural}} realization, which is
not contained within the set of realizations parametrized by the ``gauge''
function $\; \varphi, \;$ but which was shown to have some preferable
properties that have enabled us to reduce the original field theory on $\;
\kappa$-deformed space to the nonlocal, relativistically invariant,
scalar field theory on Minkowski space-time. The result of this reduction has also been
shown to not depend on the realization/ordering taken for $\;\kappa$-Minkowski
space. Thus to conclude, there is one particular noncommutative star
product on the algebra of functions, namely the one that corresponds
to the {\em{natural}} realization. Once this star product is fixed,
the realization of NC space alone is not important,
i.e. changing the realizations of 4-momenta and coordinates used to
define the $\;\kappa$-Minkowski space does not influence the final
result for the action.   
The line of consideration which is carried out here may appear useful
in 2+1 dimensional models of quantum gravity where the corresponding
Lie algebra is $\; SU(2) \;$ or $\; SU(1,1),$ \cite{39}.
Another possible route for future investigations is to consider other
noncommutative space-times. We believe that the line of analysis
presented here is applicable to other noncommutative spacetimes,
particularly those of Lie-algebra type.

{\bf Acknowledgment}\\
We thank T.R.Govindarajan for reading the manuscript and for helpful comments.
This work was supported by the Ministry of Science and Technology of the Republic of Croatia under 
contract No. 098-0000000-2865. 
 This work was done within the framework of
the Indo-Croatian Joint Programme of Cooperation in Science and Technology
sponsored by the Department of Science and Technology, India, and
the Ministry of Science, Education and Sports, Republic of Croatia.



\setcounter{equation}{0}
\setcounter{section}{0}
\renewcommand{\theequation}{A. \arabic{equation}}
\renewcommand{\thesection}{Appendix A}
\section{Derivation of some useful identities I}
\vskip 5mm

Here, we shall derive some usefull identities that are used in
calculations taking place in sections 2 and 3, where the T-operator
and the star product
corresponding to $ \; \varphi-$realization of $\kappa$-noncommutative
space were found.
First we prove the following identity:
 \begin{equation} \label{A1}
  :N^{k}::N^{l}: = \sum_{r=0}^{ min (k,l)} r! {k \choose r}
  {l \choose r} :N^{k + l - r}:,
 \end{equation} 
where $ \; N \equiv \sum_{i = 1}^{N - 1} x_{i} {\partial}_{i}  \; $ is the
number operator and the symbol :: has the usual meaning of normal
ordering with all $ \; x_{i} \; $ coming to the left with respect to all
$ \; {\partial}_{i}. $ The variables $ \; x_{i} \; $ and
$ \; {\partial}_{i}  \; $ satisfy the ordinary Heisenberg algebra
\begin{equation} 
  [{\partial}_{i}, x_{j}] = {\delta}_{ij},  \quad   i,j = 1,..., N-1.
\end{equation} 
The proof of (\ref{A1}) will be done by induction. For the base of
induction we have $ \; :N::N: =  x_{i} {\partial}_{i} +  x_{i}x_{j}
{\partial}_{i} {\partial}_{j} = :N: + :N^{2}: \; $ and this is exactly what we get
from (\ref{A1}) by taking $ \; l = k = 1.  \; $ For the step of
induction, we assume that (\ref{A1})holds for some generic $ \; l, k\; $ and
consider the expression $ \; :N^{k}::N^{l+1}: \; $. It can be
transformed by using the identity
$$
 :N^{k}: x_{i} = x_{i} :N^{k}: + k x_{i} :N^{k-1}:
$$
and with the  help of this identity we proceed as follows:
$$
  :N^{k}::N^{l+1}: = \sum_{i=1}^{ N-1} \left( x_{i} :N^{k}: + k x_{i}
  :N^{k-1}: \right ) :N^{k}: {\partial}_{i}
$$ 
$$
 = \sum_{i=1}^{N-1} x_{i}  \sum_{r=0}^{ min (k,l)} r! {k \choose r}
  {l \choose r} :N^{k + l - r}: {\partial}_{i} +
 k \sum_{i=1}^{N-1} x_{i} \sum_{r=0}^{ min (k-1,l)} r! {k-1 \choose r}
  {l \choose r} :N^{k + l - r - 1}: {\partial}_{i}
$$
\begin{equation} \label{A2}
  =  \sum_{r=0}^{ min (k,l)} r! {k \choose r}
  {l \choose r} :N^{k + l + l - r}:  +
 k  \sum_{r=0}^{ min (k-1,l)} r! {k-1 \choose r}
  {l \choose r} :N^{k + l - r }:.
\end{equation} 
If we take  $ \; k < l \; $ and change the summation index,  after
rearranging the sums, we get
$$
    \sum_{r=0}^{ k} r! {k \choose r}
  {l \choose r} :N^{k + l + l - r}:  +
   \sum_{r=1}^{k} k (r - 1)! {k-1 \choose r - 1}
  {l \choose r-1} :N^{k + l + 1 - r }:
$$
\begin{equation} \label{A3}
  = :N^{k + l + l}:  +
   \sum_{r=1}^{k}  r! {k \choose r}
   \bigg [ {l \choose r} + {l \choose r-1}  \bigg ] :N^{k + l + 1 - r }:
 = \sum_{r=0}^{min(k,l+1)}  r! {k \choose r}
    {l+1 \choose r}  :N^{k + l + 1 - r }:,
\end{equation}
which is the identity (\ref{A1}) for  $ \; l \rightarrow l+1. \; $
 The analogous procedure can be carried out for $ \; l < k \; $
and also for the expression $ \; :N^{k+1}::N^{l}: \; $ leading to the
 same conclusion, which completes the proof.

With the help of the identity (\ref{A1}) we can now prove the following
important relation:
\begin{equation} \label{B1}
  :e^{NA}::e^{NB}: =  :e^{N(A+B+AB)}:,
\end{equation} 
which is the identity for the normally ordered exponentials and
$ \; A,B \; $ are arbitrary operators commuting with the number
operator $ \; N. $  We proceed
as follows:
$$
 :e^{NA}::e^{NB}: = \sum_{l = 0}^{\infty} \sum_{k = 0}^{\infty}
  \frac{1}{l!k!} :N^{l}::N^{k}: A^{l} B^{k}
 = \sum_{l = 0}^{\infty} \sum_{k = 0}^{\infty} \frac{1}{l!k!}
   \sum_{r=0}^{ min (k,l)} r! {l \choose r}
  {k \choose r} :N^{l + k - r}: A^{l} B^{k}
$$
\begin{equation} \label{B2}
  = \sum_{s = 0}^{\infty} \sum_{k = 0}^{s} 
   \sum_{r=0}^{ min (k,s-k)} \frac{1}{(s-k-r)!} \frac{1}{r! (k-r)!} :N^{s-r}: A^{s-k} B^{k},
\end{equation} 
where in the last line we have skipped to the new summation variable
$ \; s = k + l. \; $ To accomodate for the upper bound of the 3rd sum
in the last line of (\ref{B2}),
we split the above expression into two parts
$$
 :e^{NA}::e^{NB}: = 
  \sum_{s = 0}^{\infty} \bigg(
 \sum_{k = 0}^{[\frac{s}{2}]} \sum_{r = 0}^{k}
 \frac{1}{(s-k-r)!} \frac{1}{r! (k-r)!} :N^{s-r}: A^{s-k} B^{k} 
 + \sum_{k = [\frac{s}{2}] + 1 }^{s} \sum_{r = 0}^{s-k}
 \frac{1}{(s-k-r)!} \frac{1}{r! (k-r)!} :N^{s-r}: A^{s-k} B^{k} \bigg)
$$
$$
 = \sum_{s = 0}^{\infty} \bigg( \sum_{r = 0}^{[\frac{s}{2}]} \sum_{k = r}^{[\frac{s}{2}]}
 \frac{1}{(s-k-r)!} \frac{1}{r! (k-r)!} :N^{s-r}: A^{s-k} B^{k} 
 + \sum_{r = 0 }^{s - [\frac{s}{2}] - 1} \sum_{k = [\frac{s}{2}] + 1}^{s-r}
 \frac{1}{(s-k-r)!} \frac{1}{r! (k-r)!} :N^{s-r}: A^{s-k} B^{k} \bigg)
$$
\begin{equation} \label{B3}
  = \sum_{s = 0}^{\infty} \sum_{r = 0 }^{ [\frac{s}{2}]} \sum_{k = r}^{s-r}
 \frac{1}{(s-k-r)!} \frac{1}{r! (k-r)!} :N^{s-r}: A^{s-k} B^{k},
\end{equation} 
with the symbol $ [\;]$ denoting the lowest integer.
Under the agreement that the terms with negative integers in factorials 
 drop out, the region of summation in (\ref{B3}) can be
extended to include $ \; s  \; $ as the upper bound in the last two
sums of (\ref{B3}).
 The expression (\ref{B3}) now becomes 
$$
   \sum_{s = 0}^{\infty} \sum_{r = 0 }^{ s} \sum_{k = r}^{s}
 \frac{1}{(s-k-r)!} \frac{1}{r! (k-r)!} :N^{s-r}: A^{s-k} B^{k} 
 =  \sum_{m = 0}^{\infty} \frac{:N^{m}:}{m!} \sum_{r = 0 }^{ \infty} \sum_{k = r}^{m+r}
 \frac{1}{(m-k)!} \frac{m!}{r! (k-r)!} A^{m+r-k} B^{k}
$$
\begin{equation} \label{B4}
 = \sum_{m = 0}^{\infty} \frac{:N^{m}:}{m!} \sum_{r = 0 }^{ \infty} \sum_{a = 0}^{m}
 \frac{1}{(m-r-a)!} \frac{m!}{r! a!} A^{m-r-a} B^{a} {(AB)}^{r}.
\end{equation}
Here, in the first step we have made the change of the summation
variable as $ \; s =r+m  \; $ and then, in the second step (the second line
of (\ref{B4})) we skip to the summation variable $ \; a = k - r.  \; $
Note that in the expressions (\ref{B4}) it is understood that the terms with
factorials whose arguments are negative drop out. Making use of this
fact, the upper bound of the $ \; r$-summation in (\ref{B4})
can be reduced to $ \; m, \; $ (for $ \; r > m, \; $ the corresponding
terms drop out anyway) which finally gives
$$
 :e^{NA}::e^{NB}:
 = \sum_{m = 0}^{\infty} \frac{:N^{m}:}{m!} \sum_{r = 0 }^{m} \sum_{a = 0}^{m}
 \frac{m!}{(m-r-a)! r! a!}  A^{m-r-a} B^{a} {(AB)}^{r}
$$
\begin{equation} \label{B5}
 = \sum_{m = 0}^{\infty} \frac{:N^{m}:}{m!} {(A+B+AB)}^{m} = :e^{N(A+B+AB)}:,
\end{equation}
where, in the transition between the first and the second line,
 we have utilized the multinomial expansion
\begin{equation} \label{B6}
  {(a+b+c)}^{n} = \sum_{n_{1} = 0}^{n}
 \sum_{\substack{ n_{2} = 0 \\ n_{1}+n_{2}+n_{3}=n }}^{n} \sum_{n_{3} = 0}^{n}
            \frac{n!}{n_{1}! n_{2}! n_{3}!} a^{n_{1}} b^{n_{2}} c^{n_{3}}.
\end{equation} 

Finally, it remains to prove the relation
\begin{equation} \label{C1}
  N^{k} = \sum_{l = 0}^{k} \sum_{p = 0}^{l} \frac{ {(-1)}^{l-p}
  p^{k}}{p! (l-p)!}  :N^{l}: \equiv \sum_{l = 0}^{k} c_{l,k} :N^{l}:,
\end{equation}
It can be again done by induction. From (\ref{C1}) we see that the coefficients 
$ \; c_{l,k} \; $ are defined as
\begin{equation} \label{C2}
  c_{l,k} = \sum_{p = 0}^{l} \frac{ {(-1)}^{l-p}  p^{k}}{p! (l-p)!}.  
\end{equation}
 While the base of the
induction is trivial, for the step of induction we have
\begin{equation} \label{C3}
  N^{k+1} = N \cdot N^{k} = \sum_{l = 0}^{k} c_{l,k} :N::N^{l}:
   = \sum_{l = 0}^{k} c_{l,k} \bigg [ :N^{l+1}: + l:N^{l}: \bigg ],
\end{equation} 
where use has been made of the identity (\ref{A1}). In order to
complete the step of induction, the expression (\ref{C3}) should
be equal to $ \; \sum_{l = 0}^{k+1} c_{l,k+1} :N^{l}:,  \; $ which implies
that the following recursive relations have to be satisfied:
$$
  c_{k+1,k+1} = c_{k,k} = c_{0,0} = 1, \qquad c_{0,k} = 0  \;\;\;\;
  \mbox{for} \;\;\;\; k \neq 0,
$$
\begin{equation} \label{C4}
 c_{l,k+1} = c_{l-1,k} + l \; c_{l,k}, \qquad k > 0.
\end{equation} 
The coefficients (\ref{C2}) are easily shown to satisfy the recursions
(\ref{C4}) and this completes the proof of the relation (\ref{C1}).

Now, (\ref{C1}) can be inverted to obtain
\begin{equation} \label{C5}
   :N^{k}:  = k! {N \choose k}.
\end{equation}
Using the above identity, we have for an arbitrary operator
 $ \; f, \; $ commuting with the number operator $ \; N $ 
\begin{equation} \label{C6}
   :e^{Nf}:  =  \sum_{k = 0}^{\infty} \frac{1}{ k!} :N^{k}: f^{k} =
  \sum_{k = 0}^{\infty} {N \choose k} f^{k} = {(1 + f)}^{N} =
 e^{N \ln(1 + f)} = e^{N F},
\end{equation}
where $ \; f = e^{F} - 1. $ That is, we have
\begin{equation} \label{C7}
   :e^{N(e^{ F} - 1)}:  = e^{N F}.
\end{equation}
 


\setcounter{equation}{0}
\setcounter{section}{0}
\renewcommand{\theequation}{B. \arabic{equation}}
\renewcommand{\thesection}{Appendix B}
\section{Derivation of some useful identities II}
\vskip 5mm

In this section we prove that the relation 
\begin{equation} \label{F1}
 \varphi (A_{w}) \lim_{\substack{u \rightarrow w  \\ t \rightarrow w }}
  f(u)g(t) = \lim_{\substack{u \rightarrow w  \\ t \rightarrow w }} \varphi (A_{u}+A_{t}) f(u)g(t)  
\end{equation}
holds.
We begin with
$$
 \varphi (A_{w}) \lim_{\substack{u \rightarrow w  \\ t \rightarrow w }}
  f(u)g(t) = \sum_{k = 0}^{\infty} \frac{1}{k!} {\bigg (
 \frac{\partial^{k} \varphi}{\partial A_{w}^{k}} \bigg )}_{A_{w} = 0} {(ia)}^{k}
   \frac{\partial^{k} }{\partial w_{n}^{k}} (f(w)g(w))
$$
\begin{equation} \label{F2}
 = \sum_{k = 0}^{\infty} \frac{1}{k!} {\bigg (
 \frac{\partial^{k} }{\partial A_{w}^{k}} \bigg )}_{A_{w} = 0}
 {(ia)}^{k} 
  \sum_{m = 0}^{k} {k \choose m} \frac{\partial^{m} f(w) }{\partial w_{n}^{m}}
 \frac{\partial^{k-m} g(w) }{\partial w_{n}^{k-m}} = 
   \sum_{k = 0}^{\infty} \sum_{m = 0}^{k} \frac{{(ia)}^{k}}{k!}
   {\bigg (
 \frac{\partial^{k} \varphi }{\partial A_{w}^{k}} \bigg )}_{A_{w} = 0} {k \choose m}
   \frac{\partial^{m} f(w) }{\partial w_{n}^{m}}
 \frac{\partial^{k-m} g(w) }{\partial w_{n}^{k-m}}.
\end{equation}
From the other side, we have
$$
  \lim_{\substack{u \rightarrow w  \\ t \rightarrow w }}
   \varphi (A_{u}+A_{t})
  f(u)g(t) =  \lim_{\substack{u \rightarrow w  \\ t \rightarrow w }}
 \sum_{k = 0}^{\infty} \frac{1}{k!} {\bigg (
 \frac{\partial^{k} \varphi}{\partial {(A_{u}+A_{t})}^{k}} \bigg )}_{A_{u}+A_{t} = 0}
   {(A_{u}+A_{t})}^{k} f(u)g(t) 
$$
$$
  = \lim_{\substack{u \rightarrow w  \\ t \rightarrow w }}
  \sum_{k = 0}^{\infty} \frac{1}{k!} {\bigg (
 \frac{\partial^{k} \varphi}{\partial {(A_{u}+A_{t})}^{k}} \bigg )}_{A_{u}+A_{t} = 0}
  \sum_{m = 0}^{k} {k \choose m} A_{u}^{m} A_{t}^{k-m} f(u)g(t)
$$
\begin{equation} \label{F3}
 = \sum_{k = 0}^{\infty} \sum_{m = 0}^{k} \frac{{(ia)}^{k}}{k!}
   {\bigg (
 \frac{\partial^{k} \varphi }{\partial {(2A_{w})}^{k}} \bigg )}_{2A_{w} = 0} {k \choose m}
   \frac{\partial^{m} f(w) }{\partial w_{n}^{m}}
 \frac{\partial^{k-m} g(w) }{\partial w_{n}^{k-m}}
\end{equation}
and this is identical with (\ref{F2}), which completes the proof.


\end{document}